\documentclass[12pt,a4paper]{article}

\usepackage[english]{babel}
\usepackage{epsfig}
\usepackage{amsmath}
\usepackage{amssymb}
\usepackage{amsbsy}
\usepackage{amsfonts}
\usepackage{array}
\usepackage{verbatim}
\usepackage{graphicx}
\usepackage{epstopdf}
\usepackage{subfigure}
\usepackage{tabularx}
\usepackage{bm}
\usepackage{booktabs}
\usepackage[font=small]{caption}
\usepackage{stmaryrd}
\usepackage{mathtools}

\newcommand{\mS}{\mathcal{S}}

\newcommand{\bu}{\mathbf{u}}

\newcommand{\bbar}{\overline}

\newcommand{\frho}{\bbar{\rho}}

\newcommand{\nfu}{\bbar{u}}
\newcommand{\nfS}{\bbar{\mathcal{S}}}

\newcommand{\wt}{\widetilde}

\newcommand{\fu}{\wt{u}}

\newcommand{\fS}{\wt{\mathcal{S}}}
\newcommand{\fSij}{\wt{\mS}_{ij}}

\newcommand{\de}{\partial}

\newcommand{\tauij}{\tau_{ij}}

\newcommand{\bx}{\mathbf{x}}

\begin{document}

\title{\textit{A priori} tests of a novel LES approach to compressible variable density turbulence}
\author{Caterina Bassi$^{(1)}$\\
 Antonella Abb\`a $^{(2)}$, Luca Bonaventura $^{(1)}$, Lorenzo Valdettaro $^{(1)}$}
\maketitle

\begin{center}
{\small
$^{(1)}$ MOX -- Modelling and Scientific Computing, \\
Dipartimento di Matematica, Politecnico di Milano \\
Via Bonardi 9, 20133 Milano, Italy\\
{\tt caterina.bassi@polimi.it, luca.bonaventura@polimi.it, lorenzo.valdettaro@polimi.it \\}
}
{\small
$^{(2)}$ Dipartimento di Scienze e Tecnologia Aerospaziali, Politecnico di Milano \\
Via La Masa 34, 20156 Milano, Italy\\
{\tt antonella.abba@polimi.it}
}
\end{center}

\date{}

\noindent
{\bf Keywords}:  Large Eddy Simulation, compressible flows, variable density flows, density currents, Discontinuous Galerkin method

\vspace*{0.5cm}

\noindent
{\bf AMS Subject Classification}:  65M60,65Z05,76F25,76F50,76F65

\vspace*{0.5cm}

\pagebreak


\abstract{We assess the viability of a recently proposed novel approach to LES for compressible variable density flows by means of \textit{a priori} tests. The \textit{a priori} tests have been carried out filtering a two-dimensional DNS database of the classic lock-exchange benchmark. The  tests confirm that additional terms should be accounted for in subgrid scale modeling of variable density flows, with respect to the terms usually considered in the traditional approach. Several alternatives for the modeling of these terms are assessed and discussed.}

\pagebreak
 
\section{Introduction}
\label{sec:intro} \indent  

The limitations of the conventional approaches to Large Eddy Simulation (LES) of compressible, variable density flows have been recently discussed in \cite{germano:2014}, where the importance of additional contributions to the subgrid-scale terms in presence of strong density gradients is highlighted.
 Moreover, a first proposal for the modelization of these contributions is suggested.
The purpose of the present work is to carefully assess the theoretical results in \cite{germano:2014} by means of \textit{a priori} tests. In particular, the relative importance of the different contributions to the subgrid scale stresses and the validity of the modeling proposals of \cite{germano:2014} are evaluated. 

The \textit{a priori} tests have been carried out by filtering a two-dimensional Direct Numerical Simulation (DNS) database of the classic lock-exchange benchmark. 
We have chosen this test case because it has been widely investigated, both experimentally in \cite{britter:1978}, \cite{huppert:1980}, \cite{keller:1991}, \cite{shin:2004}, \cite{simpson:1997} and numerically in
\cite{birman:2005}, \cite{bonometti:2011}, \cite{constantinescu:2014}, \cite{hartel:2000},  \cite{klemp:1997}, 
\cite{ooi:2007}, \cite{ozgokmen:2007}, \cite{ozgokmen:2009}, \cite{ungarish:2005}. This test case is also particularly appealing since it concerns complex flow evolution and turbulence phenomena, with breaking internal waves and Kelvin-Helmoltz instabilities, while being specified by simple initial and boundary conditions, see the discussion in \cite{ozgokmen:2007}.
Notice that we will focus here on the  non-Boussinesq regime, which allows for strong density differences
and which has not generally been addressed in the literature.
Due to  the transient character of the test-case, however, the statistical tools usually employed for the analysis of homogeneous or steady turbulent flows are not applicable in this context. 

The numerical technique employed in the present investigation is a  Discontinuous Galerkin (DG)  discretization, see e.g. \cite{bassi:1997}, \cite{chavent:1989}, \cite{cockburn:1989}. In particular, 
a modal DG discretization is employed, along the lines discussed in detail in \cite{abba:2015}, which has
already been validated for lock-exchange simulations in \cite{bassi:2017}.
This framework allows to compute in a straightforward way the filtered quantities as a projection onto a polynomial space of lower dimension with respect to the one employed for the DNS.

In this work, we show that some terms introduced in \cite{germano:2014}, which are usually neglected in the common density weighting approach to turbulence models for compressible turbulence, are not negligible. Furthermore, we show that the modeling proposal made in \cite{germano:2014} is also partially in contrast with the \textit{a priori} tests.
Two alternative proposals for turbulence modeling in variable density compressible flows are then assessed in this work.
The first approach is based on the modellization of the  leading subgrid stress terms following the eddy viscosity hypothesis. The \textit{a priori} tests show low values for the correlations between the exact subgrid scale terms and the modeled ones, suggesting that the eddy viscosity approach may not be the best
choice. If, despite the low correlations values, the eddy viscosity approach is preferred, the \textit{a priori} tests results suggest the introduction of two different,
dynamically computed eddy viscosities $\nu_{\rm 1}$ and $\nu_{\rm \rho}.$  
The introduction of a scale similarity model for  the leading subgrid stress terms considerably improves the results in terms of correlations. The correlations associated to the proposed similarity scale model are also higher than those associated to the traditional similarity scale approach for compressible flows.  

The paper is organized as follows. Section \ref{sec:turb_models} summarizes the results in \cite{germano:2014}. Section \ref{sec:results} is devoted to the presentation of the \textit{a priori} tests results. In section \ref{alt_mod}, alternative modeling approaches to those originally
introduced in \cite{germano:2014}
are presented and assessed, while conclusions and perspectives for future developments are drawn in section \ref{sec:conclu}. 

\section{Turbulence models for compressible variable density flows}
\label{sec:turb_models}
This section summarizes the theoretical results presented in \cite{germano:2014} on LES modeling for 
variable density, compressible flows. 
We start considering the  incompressible Navier-Stokes equations. The usual approach to LES for incompressible flows consists in the application of a filter $\overline{\cdot}$ to the Navier-Stokes equations.  When filtering the convective term in the momentum equation, this leads to the appearance of the following additional subgrid scale stress tensor:
\begin{equation}
\tau(u_i,u_j) = \overline{u_i u_j} - \overline{u}_i\,\overline{u}_j.
\end{equation}
The most popular approach to model the subgrid stresses is based on the eddy 
viscosity concept and can be formulated as:
\begin{equation}
 \tau(u_i,u_j)  = - \nu_{{\rm sgs}} \nfS_{ij}, 
 \label{eq:tau_ij_incompr}
\end{equation}
where $\nfS_{ij}$ are the components of the strain rate tensor of the resolved velocity field 
$\overline{\mathbf u}$ and the subgrid viscosity $\nu_{\rm sgs}$ can be modeled, for example, using a Smagorinsky like model (\cite{smagorinsky:1965}, \cite{germano:1991}). 

The filtering of the compressible Navier-Stokes equations is  more complex than that of the incompressible equations, since the advective term in the momentum equation is represented by a third order term $\rho u_i u_j.$
Furthermore,  a second order term $\rho u_i$ represents the advective term of the continuity equation. In order to avoid the appearance of subgrid terms in the continuity equation,  Favre filtering $\wt{\cdot} $ is introduced as:
\begin{equation}
\label{favreav}
\wt{f} = \frac{\overline{\rho f}}{\frho},
\end{equation}
see e.g. the discussion in  \cite{sagaut:2009}.
The expression for the subgrid stress tensor in the momentum equation is then:
\begin{equation}
  \tauij = \frho \theta(u_i,u_j) = \bbar{\rho u_i u_j} - \frho\fu_i\fu_j.
\label{eq:theta_base}
\end{equation}
Notice that, while usually the isotropic and deviatoric parts of the subgrid stress are modeled separately,  in this section the two terms are modeled together for the sake of simplicity.
By analogy to what is done in equation \eqref{eq:tau_ij_incompr} for incompressible flows, the common approach with density weighting to the modelization of $\tau_{ij}$ is given by:
\begin{equation}
\frho \theta(u_i,u_j) =  - \frho \nu^{{\rm sgs}} \fSij^d.
\label{eq:tau_ij_compr}
\end{equation}

Some theoretical arguments on the extension of relation (\ref{eq:tau_ij_incompr}) for incompressible flows to equation (\ref{eq:tau_ij_compr}) for compressible flows can be found in  \cite{speziale:1988} and \cite{yoshizawa:1986}.
The approach followed in \cite{germano:2014} is instead quite different. The filtered values  of $\overline{\rho u_i}$ and $\overline{\rho u_i u_j}$ are expressed as follows:
\begin{subequations}
\begin{align}
&\overline{\rho u_i} = \frho \fu_i = \frho\,\nfu_i + \tau(\rho,u_i), \label{eq:frhoui}\\
&\overline{\rho u_i u_j} = \frho \fu_i \fu_j + \frho \theta(u_i,u_j) = \frho\,\nfu_i\,\nfu_j + \frho \tau(u_i,u_j) \nonumber \\
&\hspace{1.3cm} +\nfu_i\tau(\rho,u_j) + \nfu_j\tau(\rho,u_i) + \tau(\rho,u_i,u_j), 
\end{align}
\label{eq:mass_unw_rel}
\end{subequations}
where $\tau(\rho,u_i)$ and $\tau(\rho,u_i,u_j)$ are the generalized subgrid moments associated to the turbulent transport of density. 
Notice that, starting from equations (\ref{eq:mass_unw_rel}), it is possible to derive basic relations between the standard filtered quantities and the Favre filtered ones as:
\begin{subequations}
\begin{align}
&\fu_i = \nfu_i + \frac{\tau(\rho,u_i)}{\frho}, \label{eq:rel_favre_norm-a}\\
&\theta(u_i,u_j) = \tau(u_i,u_j) - \frac{\tau(\rho,u_i)\tau(\rho,u_j)}{\frho^2} + \frac{\tau(\rho, u_i, u_j)}{\frho}. \label{eq:rel_favre_norm-b}
\end{align}
\label{eq:rel_favre_norm}
\end{subequations}
As pointed out in \cite{germano:2014}, equations (\ref{eq:rel_favre_norm}) are well established in the context of Reynolds and Favre averages; the introduction of the generalized central moments allows their extension to the case of a filter operator, which does not always satisfy the property $\overline{\overline{f}} = \overline{f}$. Notice that, if we substitute the expression of the Favre filtered velocity in equation (\ref{eq:rel_favre_norm-a}) into the expression for the Favre filtered strain rate, we obtain:
\begin{multline}
\fS_{ij} = \de_j \fu_i + \de_i \fu_j = \de_j \nfu_i + \de_i \nfu_j -\frac{\tau(\rho,u_i)\de_j\frho + \tau(\rho,u_j) \de_i \frho}{\frho^2} \\
\hspace{2.8cm}+ \frac{\de_j \tau(\rho,u_i) + \de_i\tau(\rho,u_j)}{\frho}.
\end{multline}
We can then rewrite $\nfS_{ij}$ as follows:
\begin{equation}
\nfS_{ij} = \fS_{ij} + \frac{\tau(\rho,u_i)\de_j\frho + \tau(\rho,u_j) \de_i \frho}{\frho^2} -  \frac{\de_j \tau(\rho,u_i) + \de_i \tau(\rho,u_j)}{\frho}
\label{eq:nfs_fs}
\end{equation}
If now we substitute $\tau(u_i,u_j)$, modeled as in equation (\ref{eq:tau_ij_incompr}), in equation (\ref{eq:rel_favre_norm-b}) and we use equation (\ref{eq:nfs_fs}), we have: 
\begin{eqnarray}
\theta(u_i,u_j) &=& -\nu_{\rm sgs} \left[ \fS_{ij} + \frac{\tau(\rho,u_i)\de_j\frho + \tau(\rho,u_j) \de_i \frho}{\frho^2} 
\right. \nonumber \\
&-& \left. \frac{\de_j \tau(\rho,u_i) + \de_i \tau(\rho,u_j)}{\frho} \right] \nonumber\\
\hskip 0.2cm &-&\frac{\tau(\rho,u_i)\tau(\rho,u_j)}{\frho^2} + \frac{\tau(\rho, u_i, u_j)}{\frho}.
\label{eq:theta_complete}
\end{eqnarray} 
If we consider an eddy viscosity model also for the terms $\tau(\rho,u_i)$ and $\tau(\rho,u_i,u_j)$:
\begin{subequations}
\begin{align}
&\tau(\rho,u_i) = -\nu_{\rm \rho} \de_i \frho, \\
&\tau(\rho,u_i,u_j) = - \nu_{\rm \rho u} (\de_j \tau(\rho,u_i)+\de_i \tau(\rho,u_j)),
\label{eq:eddy_visc_hyp}
\end{align}
\end{subequations}
we can notice that the conventional hypothesis (\ref{eq:tau_ij_compr}) is valid if the three eddy viscosities $\nu_{\rm sgs}$, $\nu_{\rm \rho}$ and $\nu_{\rm \rho u}$ satisfy the following hypothesis:
\begin{equation*}
\nu_{\rm \rho u} = \nu_{\rm sgs}, \quad \nu_{\rm \rho} = 2\nu_{\rm sgs},
\end{equation*}
which are not generally valid.

In \cite{germano:2014}, an attempt is made to take into account some of the additional terms in equation (\ref{eq:theta_complete}). In particular, if equations (\ref{eq:eddy_visc_hyp}) together with equation (\ref{eq:tau_ij_incompr}) are assumed and the following hypothesis are considered
\begin{equation}
\nu_{\rm \rho u} = \nu_{\rm sgs}, \quad \nu_{\rm \rho} \neq 2\nu_{\rm sgs},
\label{eq:hyp_germano}
\end{equation}
$\theta(u_i,u_j)$ can be expressed as:
\begin{equation}
\theta(u_i,u_j) = -\nu_{\rm sgs}(\de_j \fu_i + \de_i \fu_j) - \frac{\nu_{\rm \rho}(\nu_{\rm\rho}-2\nu_{\rm sgs})}{\frho^2} \de_i\frho \de_j \frho. 
\end{equation}

The different terms of equations (\ref{eq:rel_favre_norm}) and (\ref{eq:theta_complete}) will be carefully estimated by means of an \textit{a priori} test, whose results are presented in the following section, in order to establish whether the hypotesis (\ref{eq:hyp_germano}) can be actually considered valid.

\section{\textit{A priori} tests results}
\label{sec:results}
\begin{figure}
\centering
\includegraphics[width=0.5\textwidth]{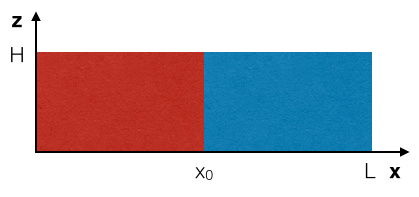}
\caption{Initial datum for the lock-exchange configuration.}
\label{fig:lock_exchange_domain}
\end{figure}

The lock-exchange configuration employed in  the \textit{a priori} tests is represented in figure \ref{fig:lock_exchange_domain}. In non dimensional units, the domain length is $L=5$ and its height is $H=1,$
while the total duration of the simulation is $T=25$. A membrane initially divides the rectangular container in two compartments (the position of the membrane is $x_0=2.5$ in the present computations). In our case, the two chambers are filled with the same fluid at different densities on the two sides of the membrane (higher density on the left and lower density on the right). Upon the removal of the membrane, the dense front moves rightward along the lower boundary, while the light front propagates leftward along the upper boundary.
The ratio between the initial densities is $\gamma_r=0.4$, the Mach number is $Ma=0.1$, while the Reynolds number is equal to $Re=2800$. 

Notice that, as previously remarked, the model equations (compressible Navier-Stokes equations with gravity),    their non dimensional
formulation and the numerical discretization are the same as presented in \cite{abba:2015} and \cite{bassi:2017},
to which we refer for a complete description of the numerical method.  Time integration has been performed with  a five stages Strong Stability Preserving Runge-Kutta method described in \cite{spiteri:2002}. 

Concerning the initial conditions, the initial density profile is given by: 
\begin{equation}
\rho_0(x) = \frac{\gamma_r+1}{2} - \frac{1-\gamma_r}{2} {\rm erf}
\left( \frac{x-x_0}{\sqrt{Re}} \right),
\label{init_dens_prof}
\end{equation}
where $x$ denotes the horizontal coordinate  (\cite{bassi:2017}, \cite{birman:2005}).
Since we are considering the compressible Navier-Stokes equations, it is necessary to specify the initial conditions also for pressure and temperature. 
The initial pressure distribution in the domain is computed assuming an hydrostatic pressure profile where the initial value at the top of the domain is imposed as in \cite{bassi:2017}.
The initial datum for temperature is derived starting from density and pressure and using the equation of state. 
Concerning the boundary conditions, the same slip boundary conditions as in \cite{bassi:2017} have been imposed.

For the space discretization, the polynomial degree  $p=7$ was employed, which entailed a  number of degrees of freedom per element equal to $N_p = (p+1)(p+2)/2=36$. The choice of the polynomial degree and of the computational grid (composed approximately of $4000$ elements) was made so as to obtain a total number of degrees of freedom similar to the one employed in \cite{ozgokmen:2007} for two-dimensional Boussinesq simulations at the same Reynolds number. 
The mesh is built starting from a structured Cartesian mesh with $N_x=104$, $N_z=20$ quadrilaterals in the $x,z$ directions. Each quadrilateral is then divided into $N_t=2$ triangular elements. The mesh is uniform in all directions and the equivalent mesh spacing in each direction, taking into account the fact that high-order polynomials are employed, is given by:
\begin{equation}
\Delta_x = \frac{L}{N_x \sqrt{N_t N_p}}, \quad \Delta_z = \frac{H}{N_z\sqrt{N_t N_p}},
\end{equation}
where $L$ and $H$ are the length and height of the computational domain, respectively.

The grid filter and the test filter, necessary in order to carry out the \textit{a priori} tests, are identified with the $L_2$ projection on the space of $\overline{p}=4$ and $\widehat{p}=2$ piecewise polynomial functions, respectively. 
The grid filter scale can be computed, for the generic element $K$, as:
\begin{equation}
\Delta(K)= \frac{\overline{\Delta}_x \overline{\Delta}_z}{N_{\overline{p}}},
\end{equation}
with $\overline{\Delta}_x = \frac{L}{N_x \sqrt{N_t N_{\overline{p}}}}$ and $\overline{\Delta}_z = \frac{H}{N_z \sqrt{N_t N_{\overline{p}}}}$. 
The test filter scale is defined analogously, with the only difference that $N_{\overline{p}}$ is substituted by the number of degrees of freedom per element corresponding to the polynomial degree associated to the test filter.

\begin{figure}
\centering
\includegraphics[width=0.7\textwidth]{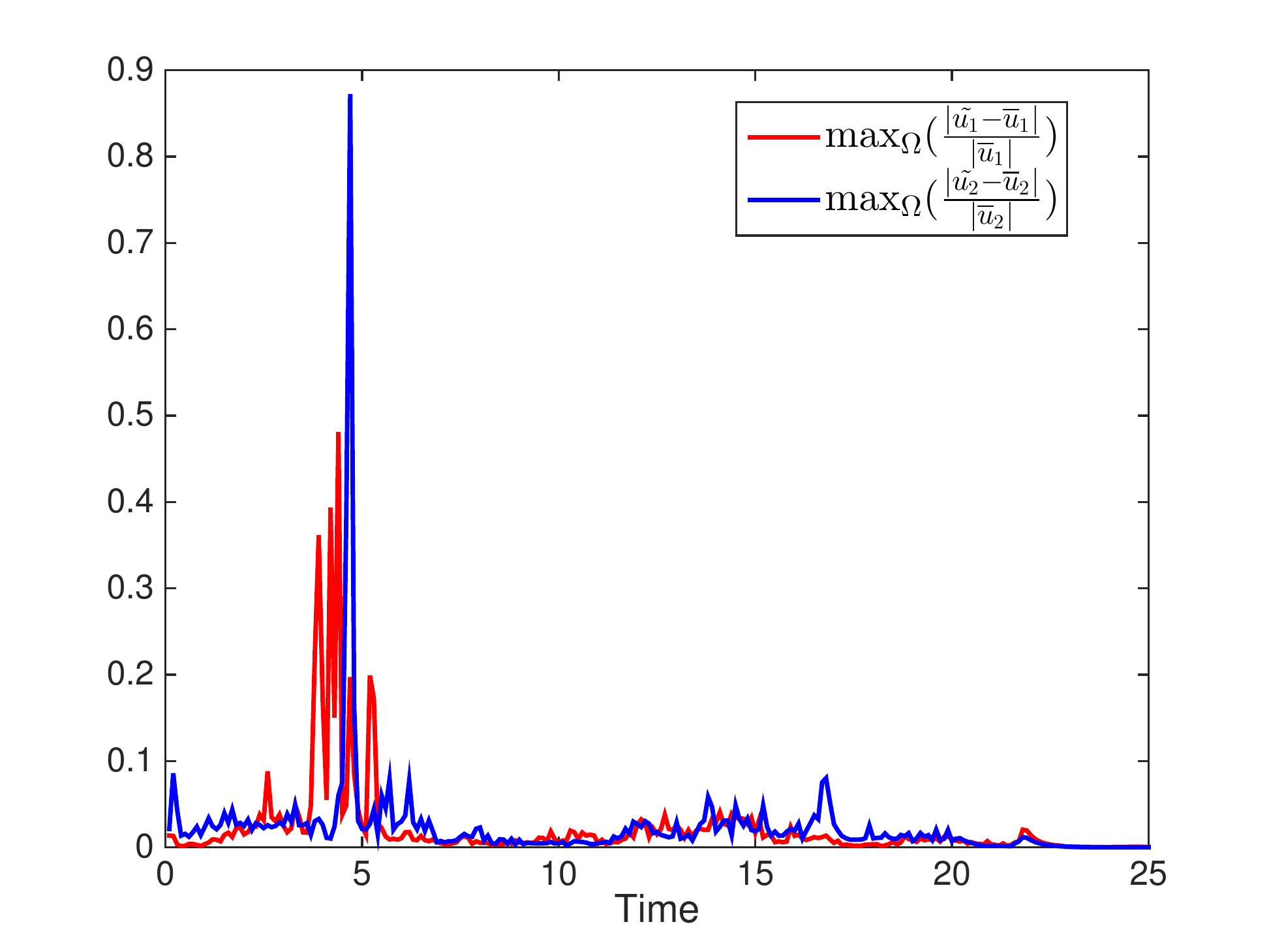}
\caption{Maximum value over the domain $\Omega$ of the relative difference between the Favre filtered velocity and the filtered velocity, as a function of time.}
\label{fig:max_vdiff}
\end{figure}

The first quantity to be evaluated in the \textit{a priori} tests is the difference, if any, between the filtered velocity and the Favre filtered velocity. The time evolution of the quantities 
$$\max_{\Omega}\left(\frac{|\fu_i-\overline{u}_i|}{|\overline{u}_i|}\right), \ \ \ \ i=1,2$$ 
is reported in figure \ref{fig:max_vdiff}. 
Here, $\Omega $ denotes the computational domain. We can notice that significant differences in the maximum values, up to $90\%$, are present. 

Having verified that significant differences between the filtered velocity and the Favre filtered velocity can arise, we consider equation (\ref{eq:rel_favre_norm-b}) and we rewrite the three contributions to $\theta(u_i,u_j)$ separately as:
\begin{subequations}
\label{eq:apriori_contrib}
\begin{align}
&\tau(u_i,u_j), \label{eq:apriori_contrib-1} \\
&b_{ij} = - \frac{\tau(\rho, u_i) \tau(\rho, u_j)}{\overline \rho^2},\label{eq:apriori_contrib-2}\\
&c_{ij} =  \frac{\tau(\rho, u_i, u_j)}{\overline \rho}. \label{eq:apriori_contrib-3}
\end{align}
\end{subequations}
The time evolution of the Frobenius norm: 
\begin{equation}
\label{eq:fr_norm_all}
\parallel \theta \parallel_{F} = \sqrt{\int_\Omega \sum_{ij} \theta(u_i,u_j)^2d\mathbf{x}}
\end{equation}
for each of  the three contributions (\ref{eq:apriori_contrib}) has been computed, together with
the norm of $\theta(u_i,u_j)$ itself.
Moreover, we have also considered the $L_2$ norm of the individual components of each tensor:
\begin{equation}
\parallel \theta(u_i,u_j) \parallel_{L_2} = \sqrt{\int_\Omega \theta(u_i,u_j)^2}d\mathbf{x}, \quad \text{for } i,j=1,\cdots,d.
\label{eq:l2norm_components}
\end{equation}
The time evolution of the maximum and minimum values 
$$\max_\Omega\theta(u_i,u_j) \ \ \ \ \min_\Omega\theta(u_i,u_j),  \ \ \ \ i,j=1,\cdots,d$$
taken by the individual components of each tensor  have also been evaluated. 
 Analogous expressions have also been computed for $\tau(u_i,u_j)$, $b_{ij}$ and $c_{ij}$.

\begin{figure}
\centering
\includegraphics[width=0.7\textwidth]{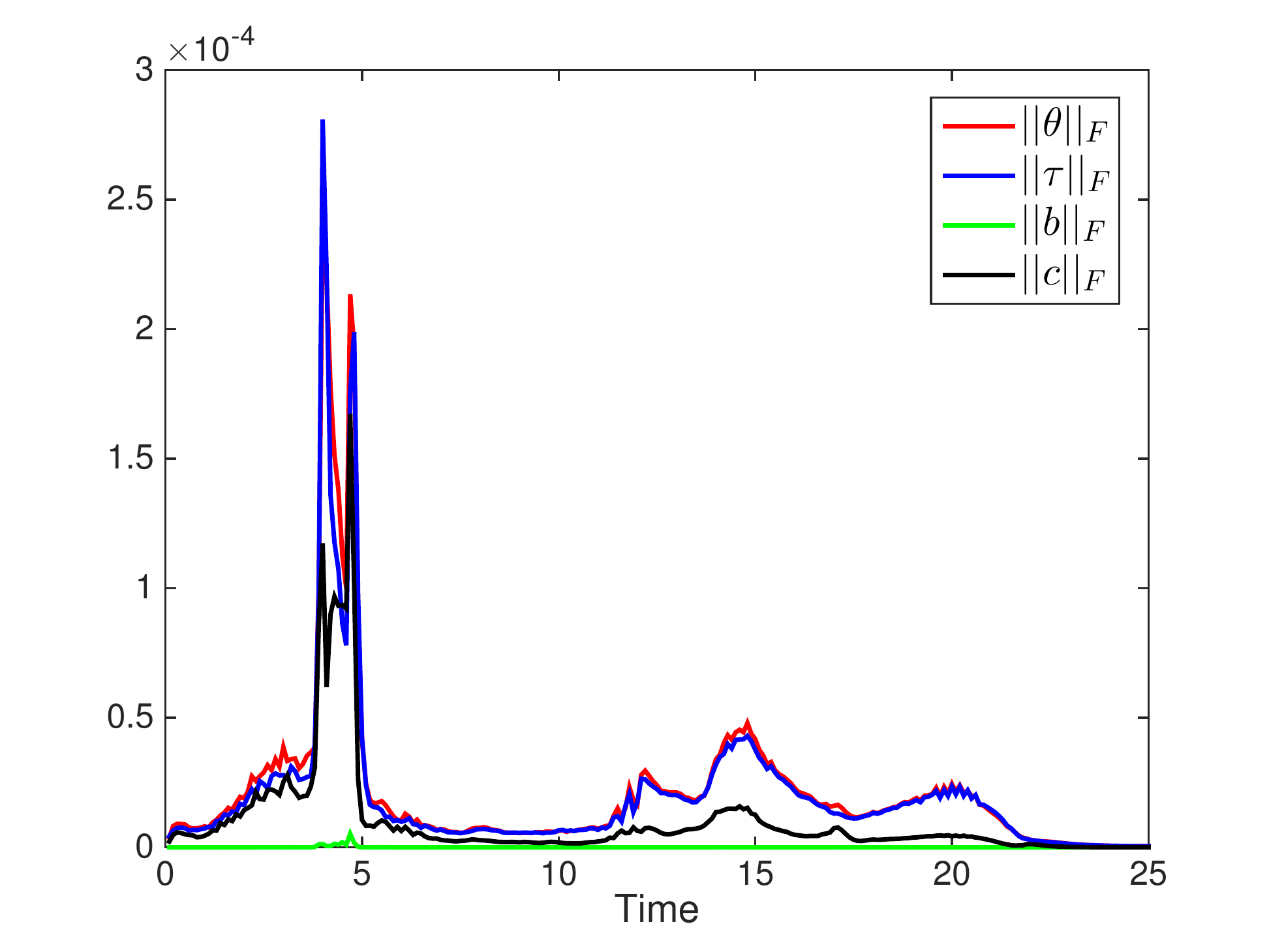}
\caption{Frobenius norm of $\theta$, $\tau$, $b$ and $c$ as a function of time.}
\label{fig:fr_norm_all}
\end{figure}

\begin{figure}[]
\centering
\begin{subfigure}[]{
      \includegraphics[width=0.56\textwidth]{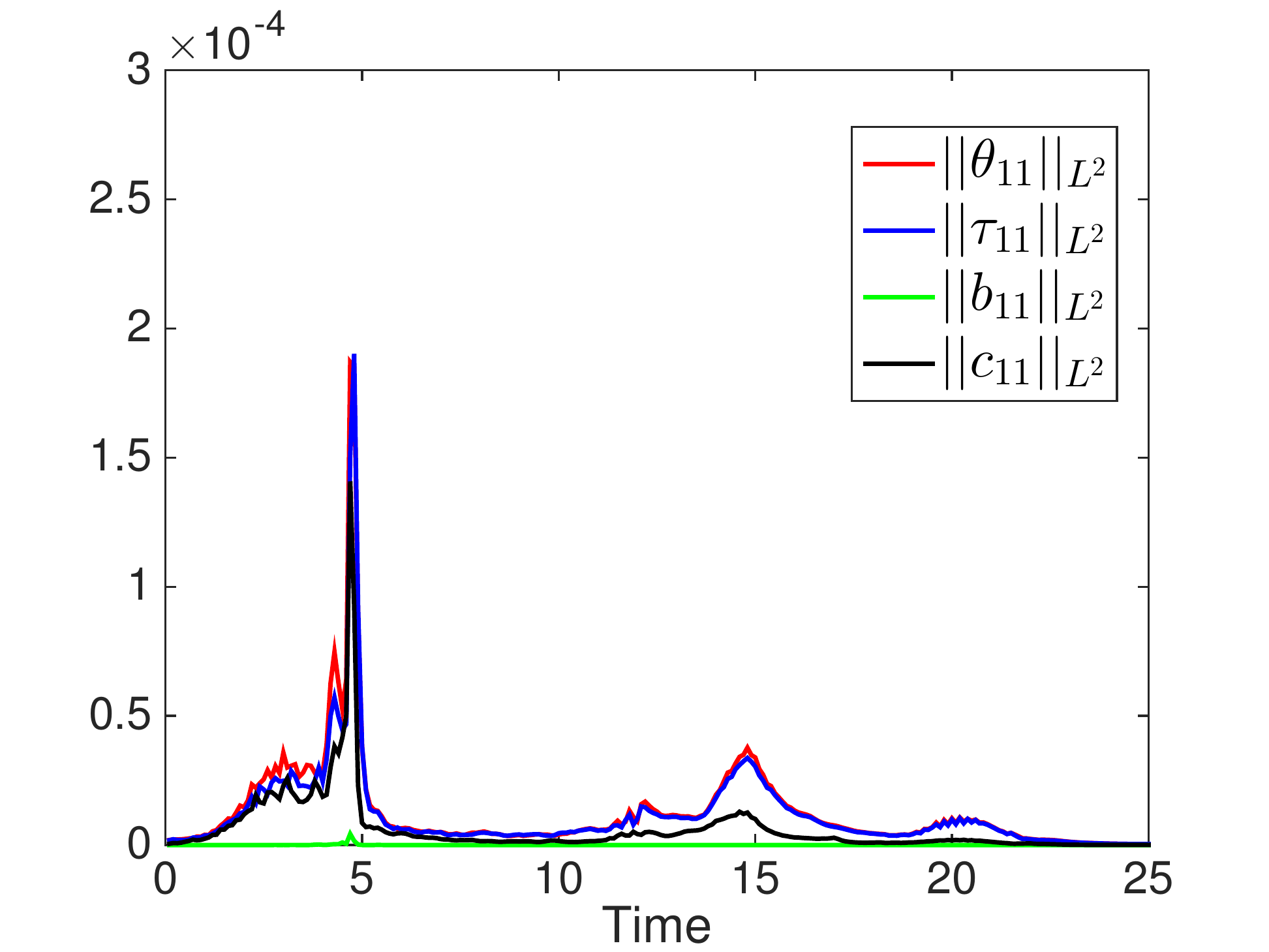}}
\end{subfigure}
\begin{subfigure}[]{
      \includegraphics[width=0.56\textwidth]{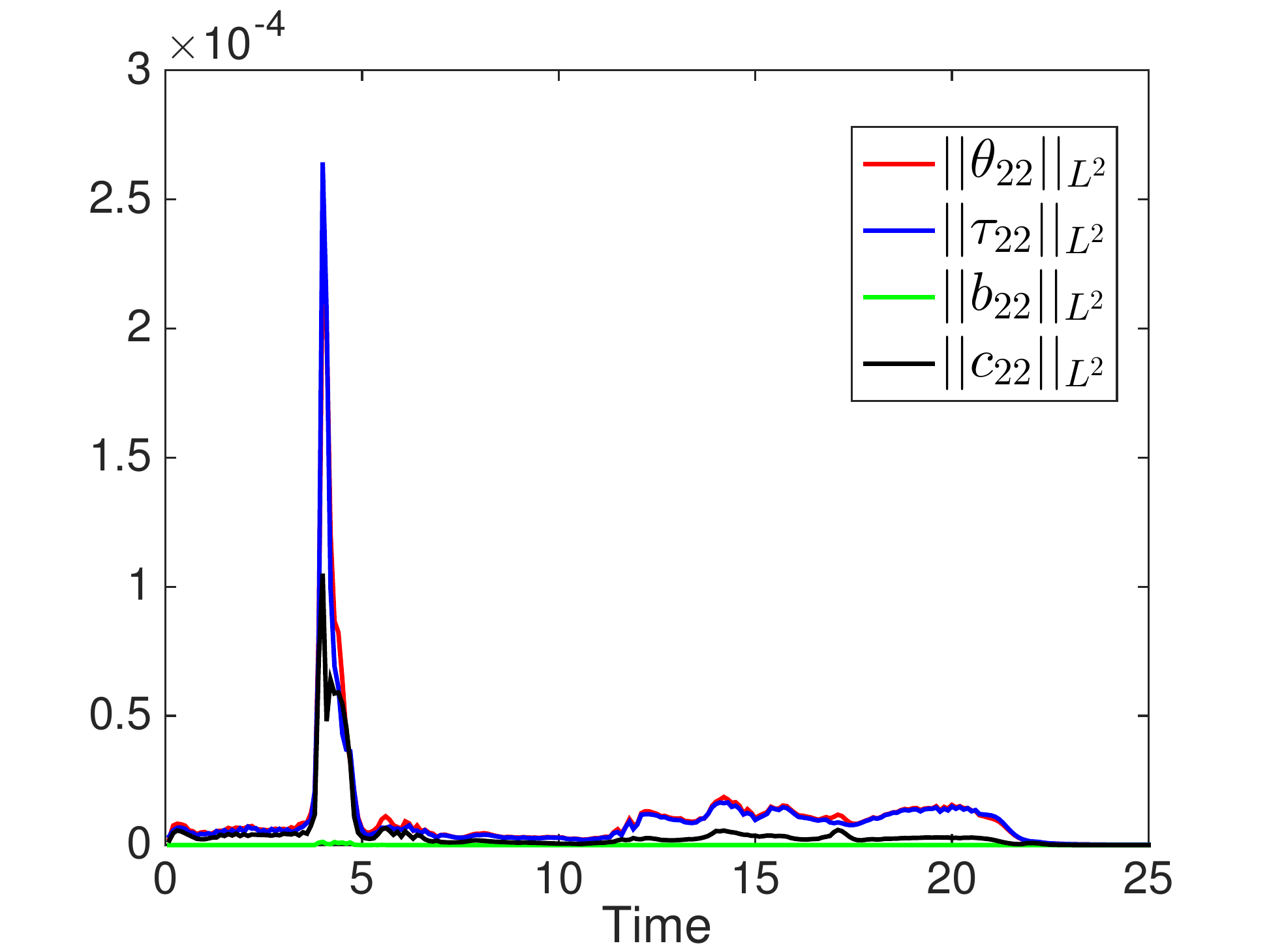}}
\end{subfigure} 
\begin{subfigure}[]{
       \includegraphics[width=0.56\textwidth]{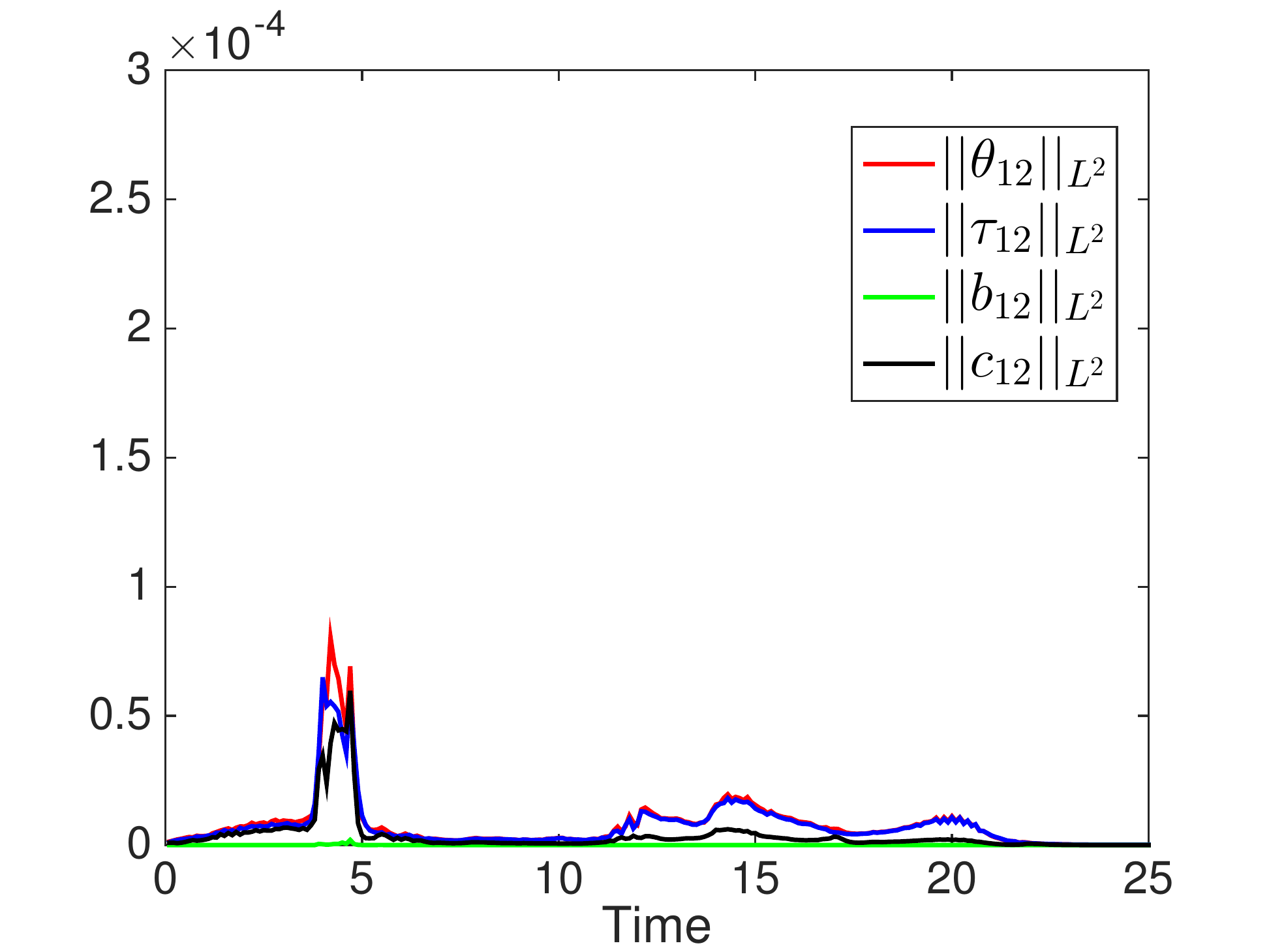}}
\end{subfigure}
\caption{$L_2$ norm of the different components of $\theta$, $\tau$, $b$ and $c$ as a function of time. (a) First diagonal component. (b) Second diagonal component. (c) Off diagonal component.}
\label{fig:l2norm_components}
\end{figure}

In figure \ref{fig:fr_norm_all}, the time evolution of the Frobenius norm (\ref{eq:fr_norm_all}) of $\theta$, $\tau$, $b$ and $c$ is shown. We can easily notice that the predominant contributions are those of $\theta$ and $\tau$. Also the norm of $c$ takes significant values, while the norm of $b$ is 3 or 4 orders of magnitude smaller. 
If we look at figure \ref{fig:l2norm_components}, we can see that the $L_2$ norms of the single components of the different tensors (see equation (\ref{eq:l2norm_components})) confirm this trend. Moreover, we can also notice that the diagonal components are slightly larger than the off-diagonal ones.

\begin{figure}[]
\centering
\begin{subfigure}[]{
      \includegraphics[width=0.56\textwidth]{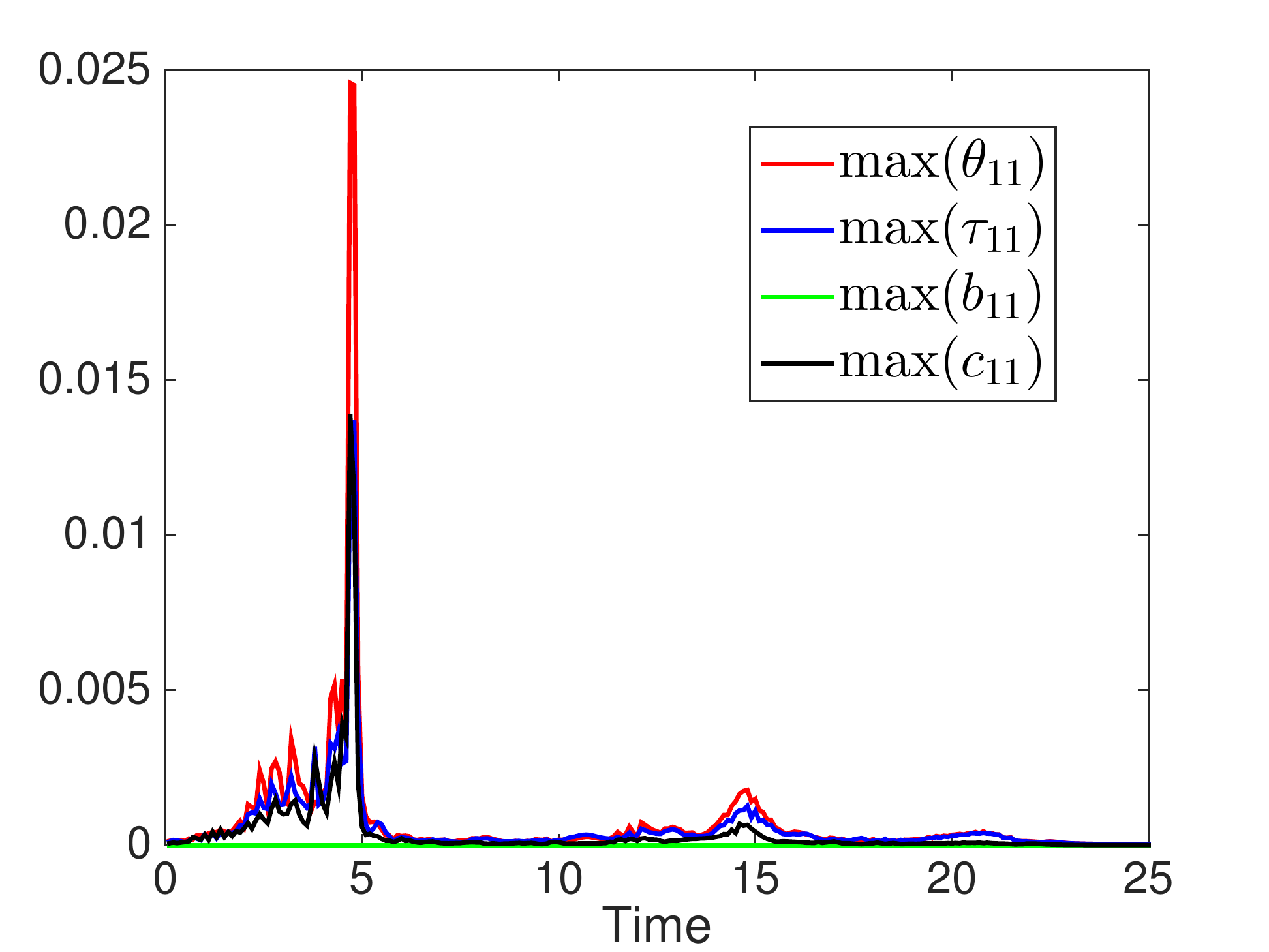}}
\end{subfigure}
\begin{subfigure}[]{
      \includegraphics[width=0.56\textwidth]{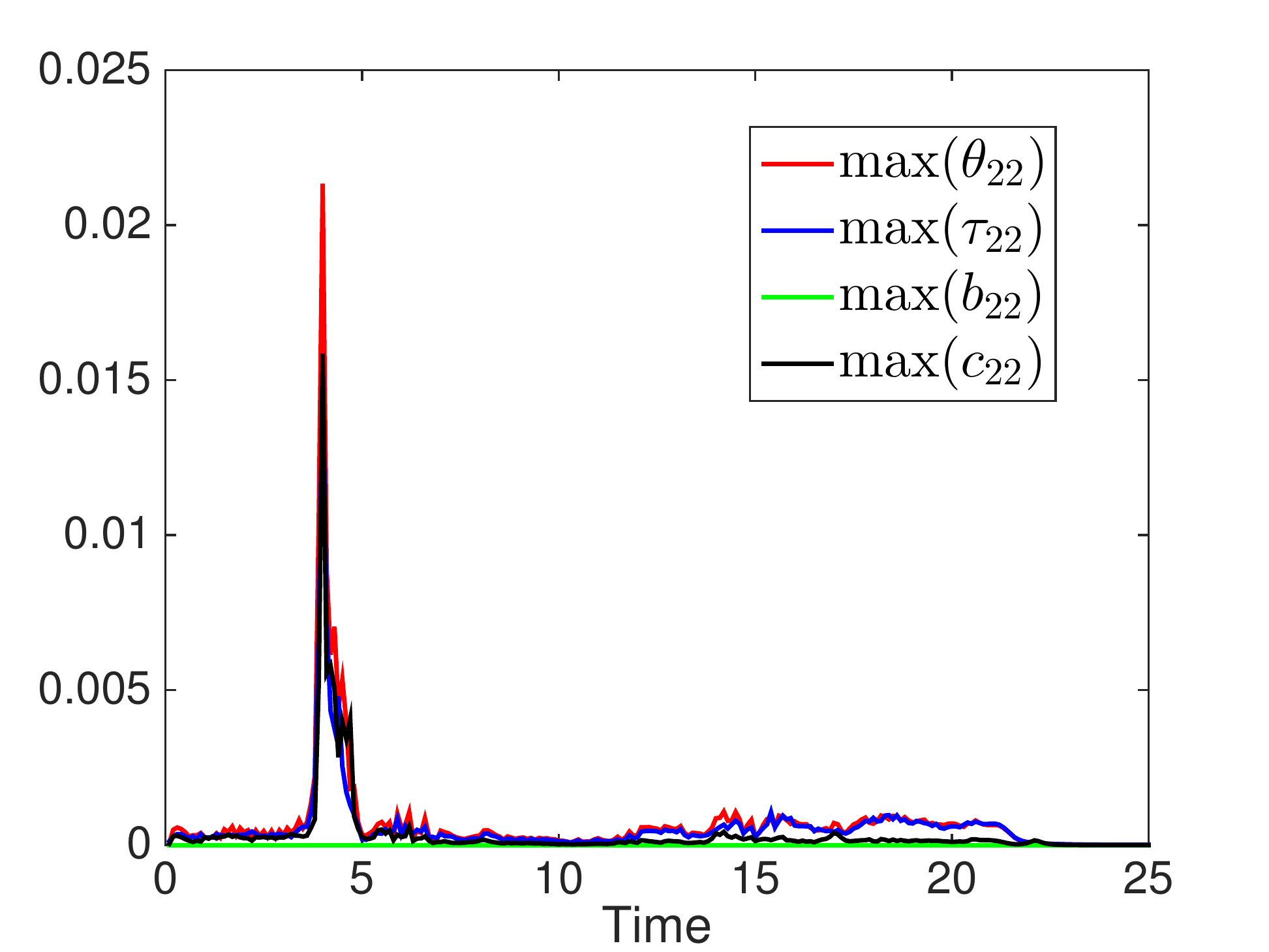}}
\end{subfigure} 
\begin{subfigure}[]{
       \includegraphics[width=0.56\textwidth]{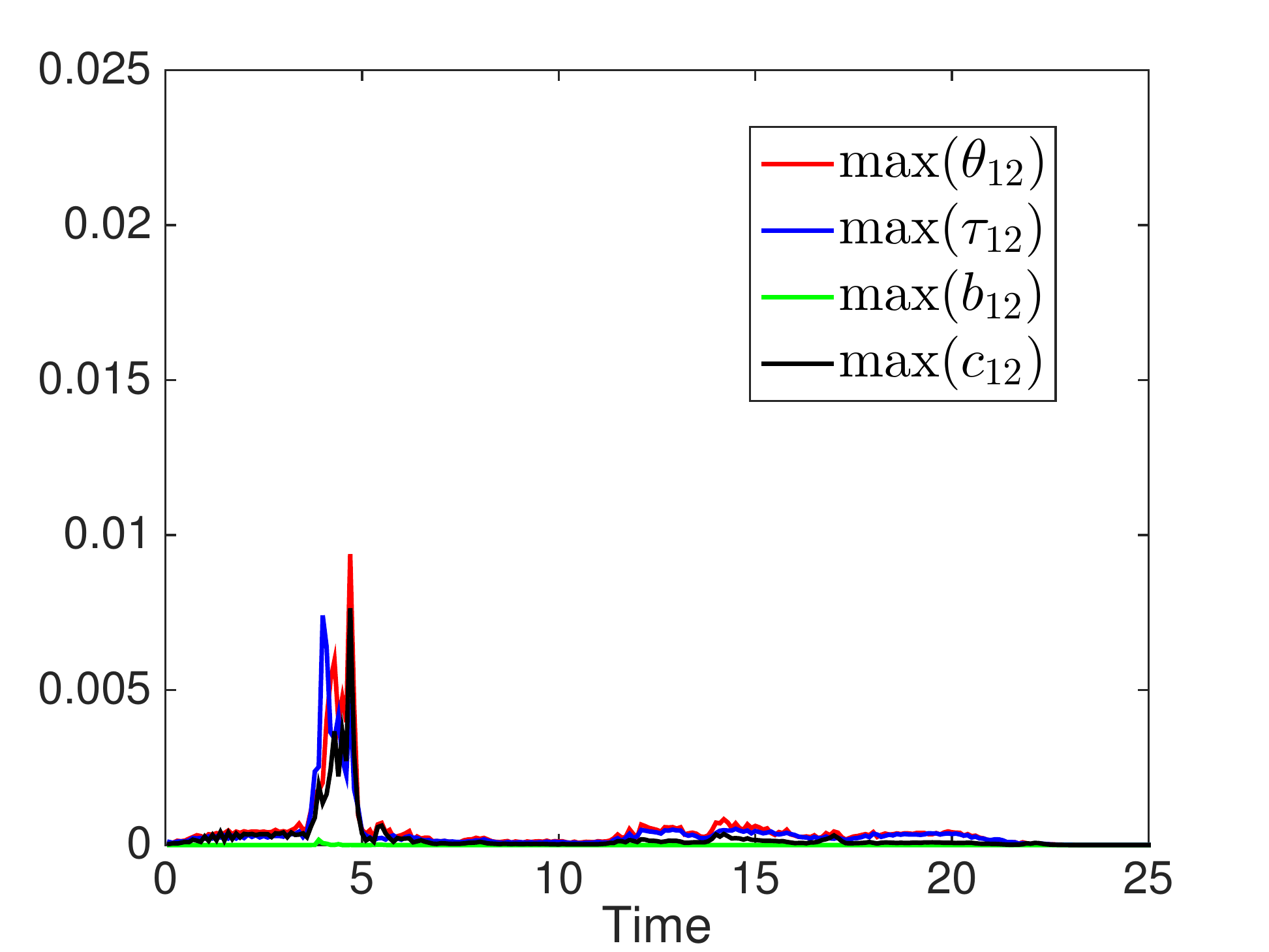}}
\end{subfigure}
\caption{Maximum value over the domain $\Omega$ of the different components of $\theta$, $\tau$, $b$ and $c$ as a function of time. (a) First diagonal component. (b) Second diagonal component. (c) Off-diagonal component.}
\label{fig:max}
\end{figure}

\begin{figure}[]
\centering
\begin{subfigure}[]{
      \includegraphics[width=0.56\textwidth]{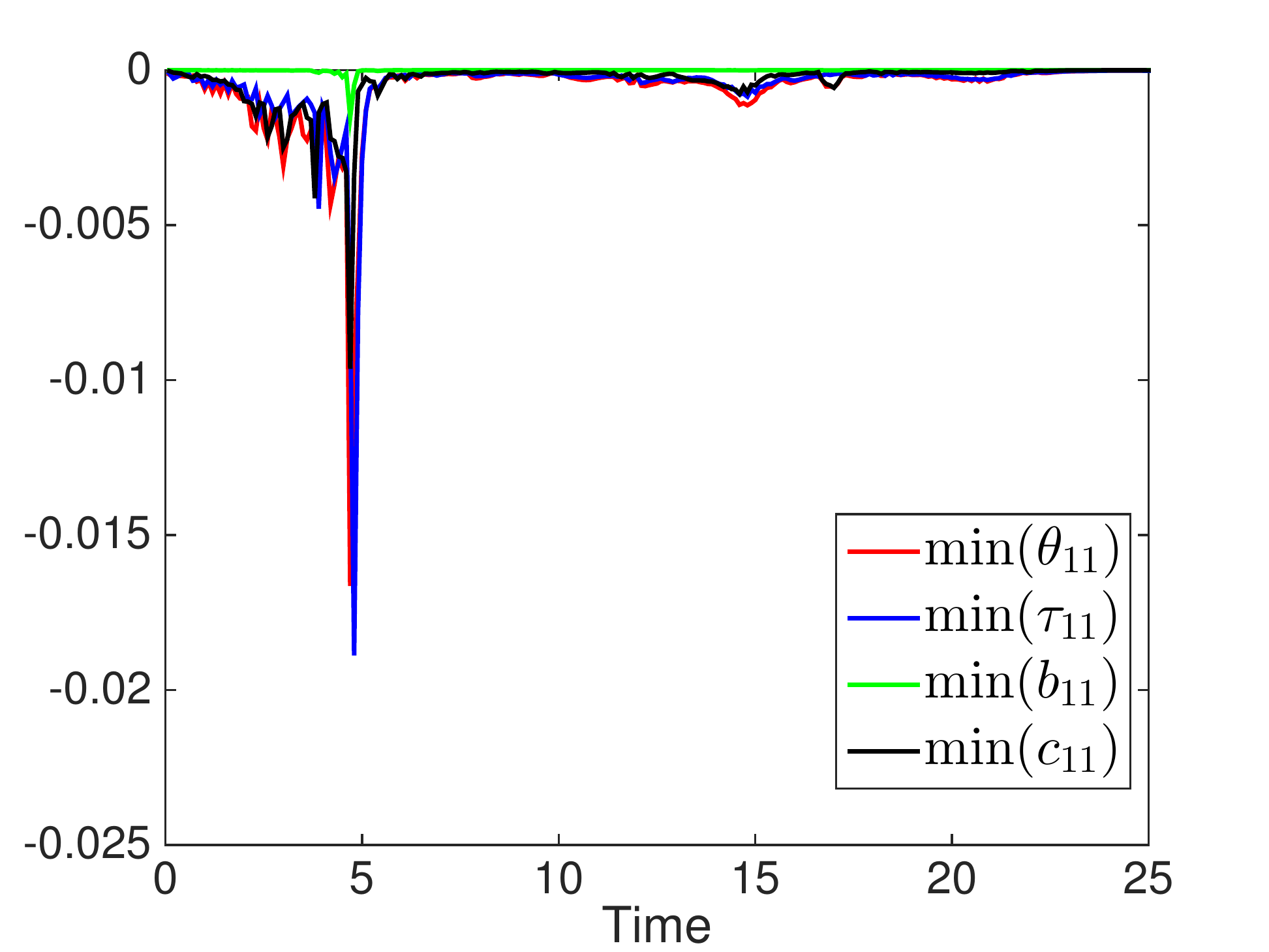}}
\end{subfigure}
\begin{subfigure}[]{
      \includegraphics[width=0.56\textwidth]{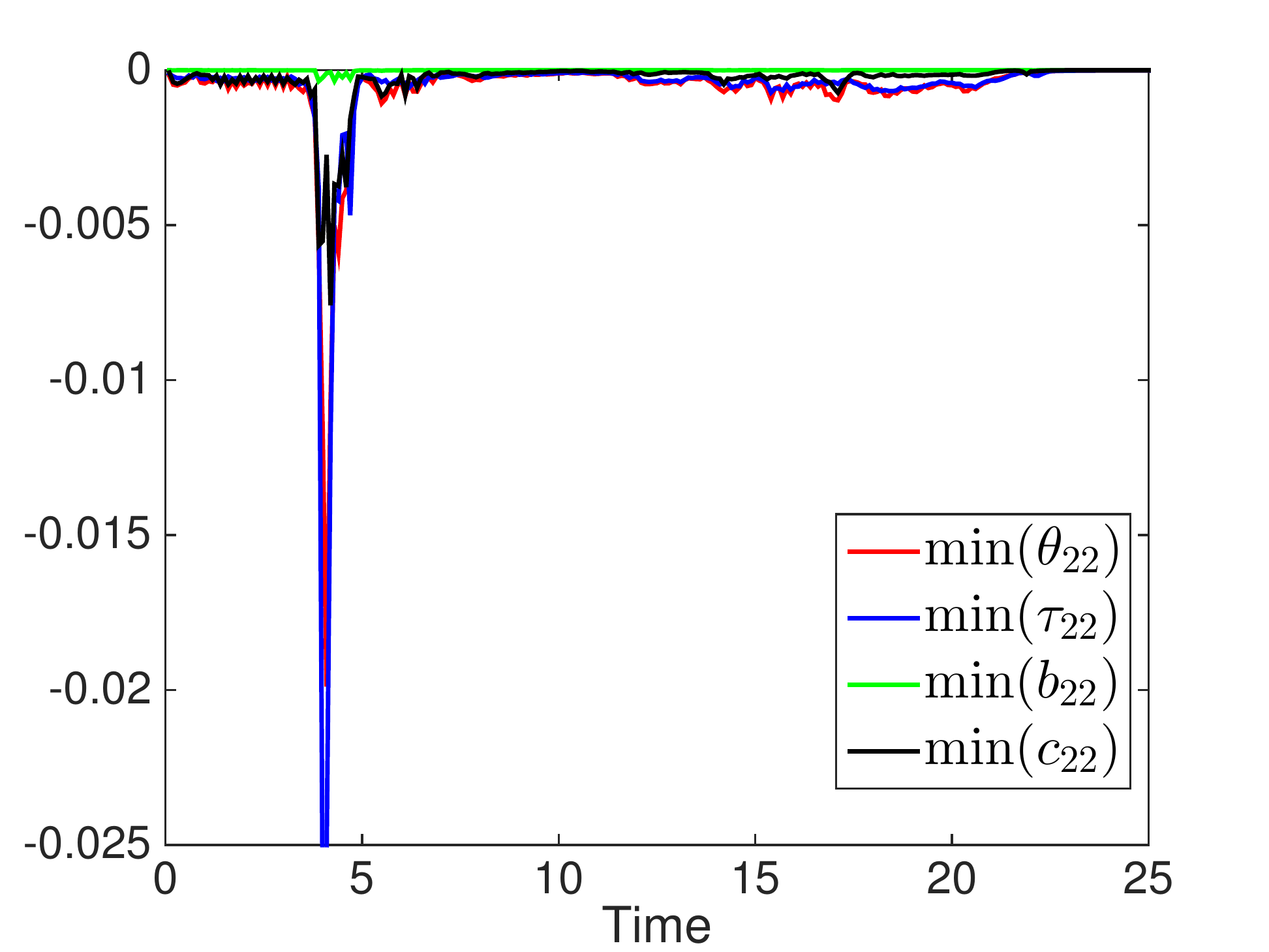}}
\end{subfigure} 
\begin{subfigure}[]{
       \includegraphics[width=0.56\textwidth]{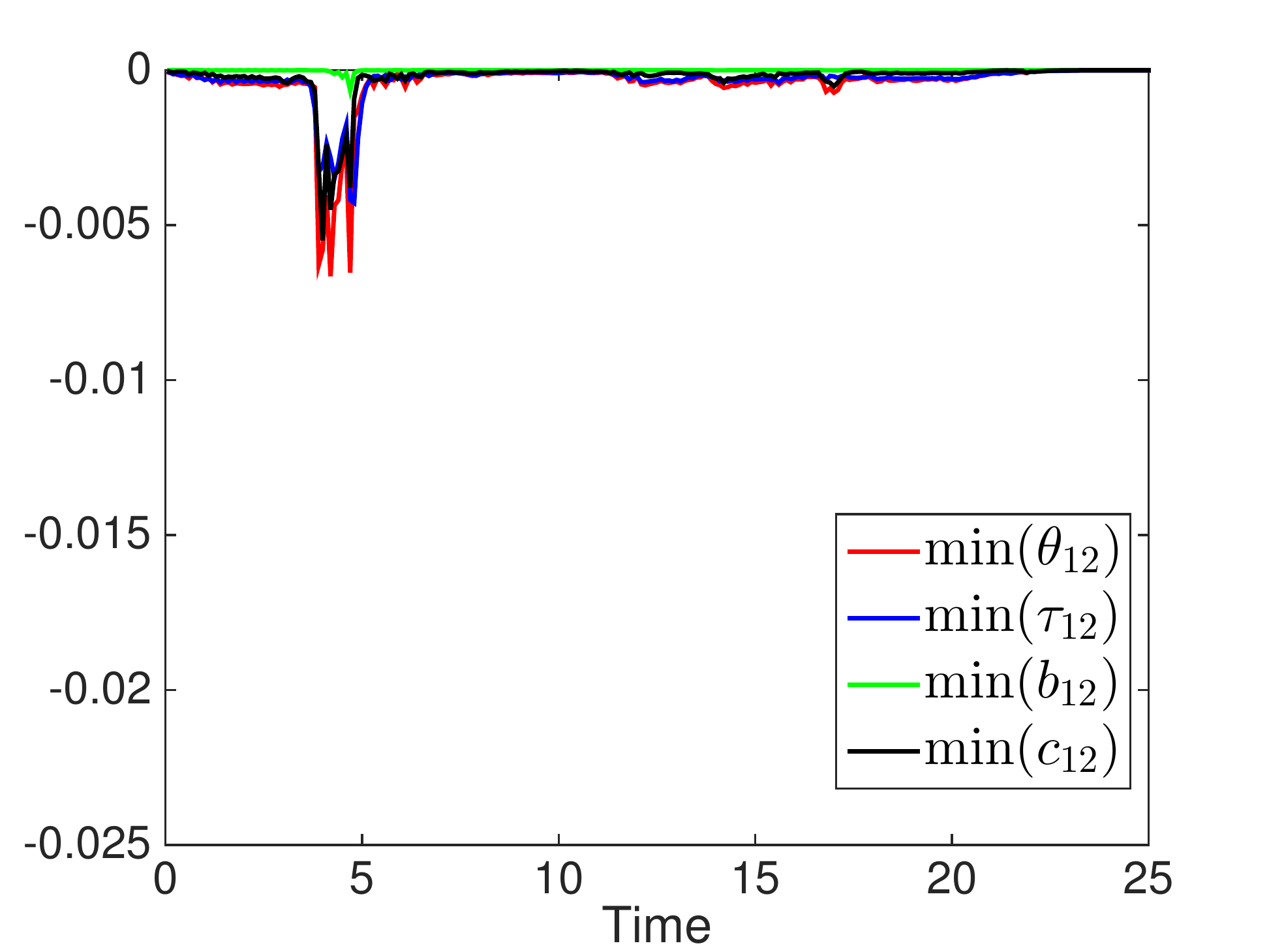}}
\end{subfigure}
\caption{Minimum value over the domain $\Omega$ of the different components of $\theta$, $\tau$, $b$ and $c$ as a function of time. (a) First diagonal component. (b) Second diagonal component. (c) Off diagonal component.}
\label{fig:min}
\end{figure}

The time evolution of the maximum (figure \ref{fig:max}) and minimum values (figure \ref{fig:min}) of the components of $\theta$, $\tau$, $b$ and $c$ is consistent with the previous results, confirming the predominance of $\tau$ and $\theta$, followed by $c$, and the fact that $b$ is far less important.

\begin{figure}
\centering
\includegraphics[width=0.7\textwidth]{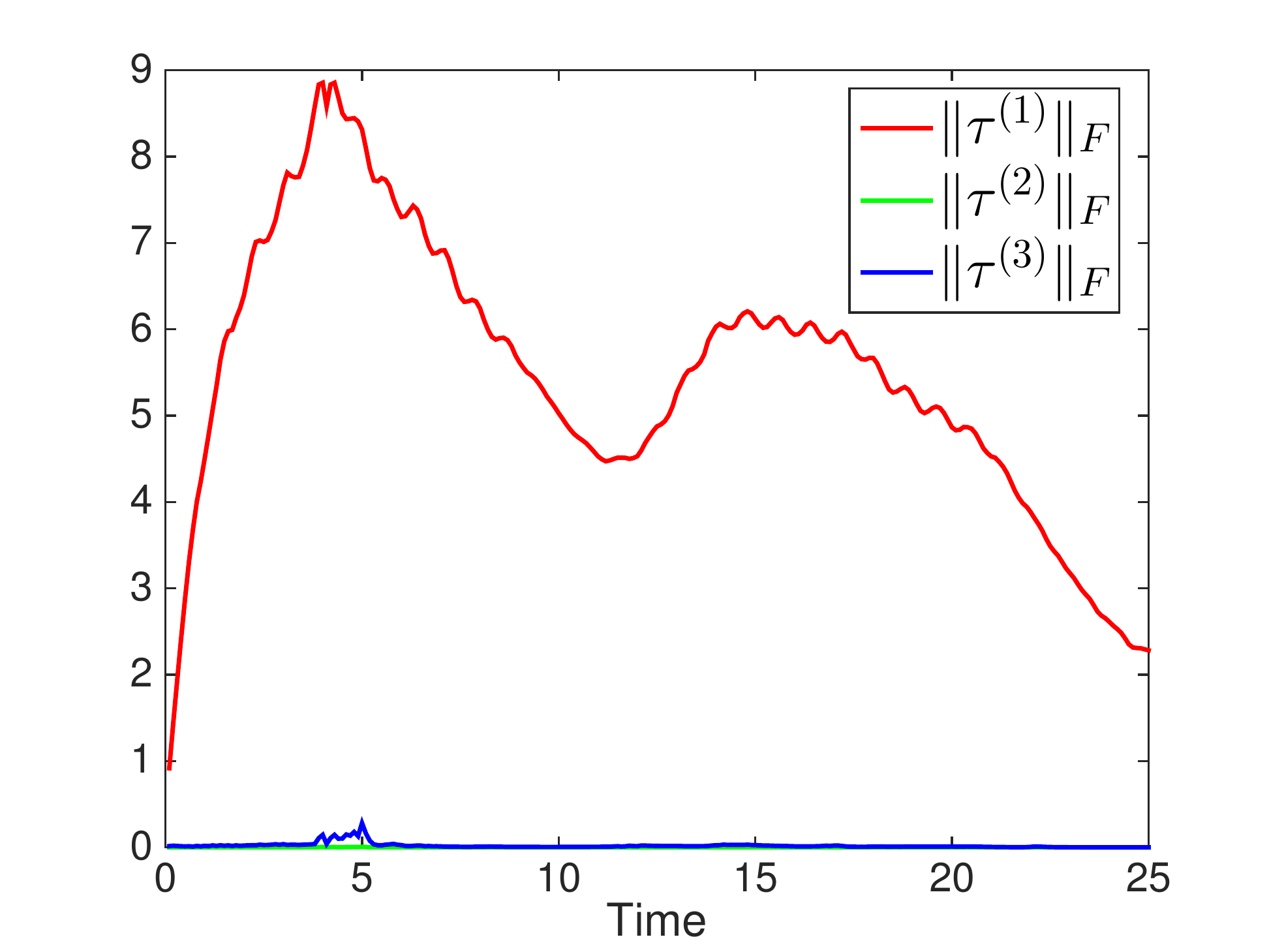}
\caption{Frobenius norm of $\tau ^{(1)}$, $\tau ^{(2)}$ and $\tau^{(3)}$ as a function of time.}
\label{fig:fr_norm_tau_contrib}
\end{figure}

\begin{figure}[]
\centering
\begin{subfigure}[]{
      \includegraphics[width=0.56\textwidth]{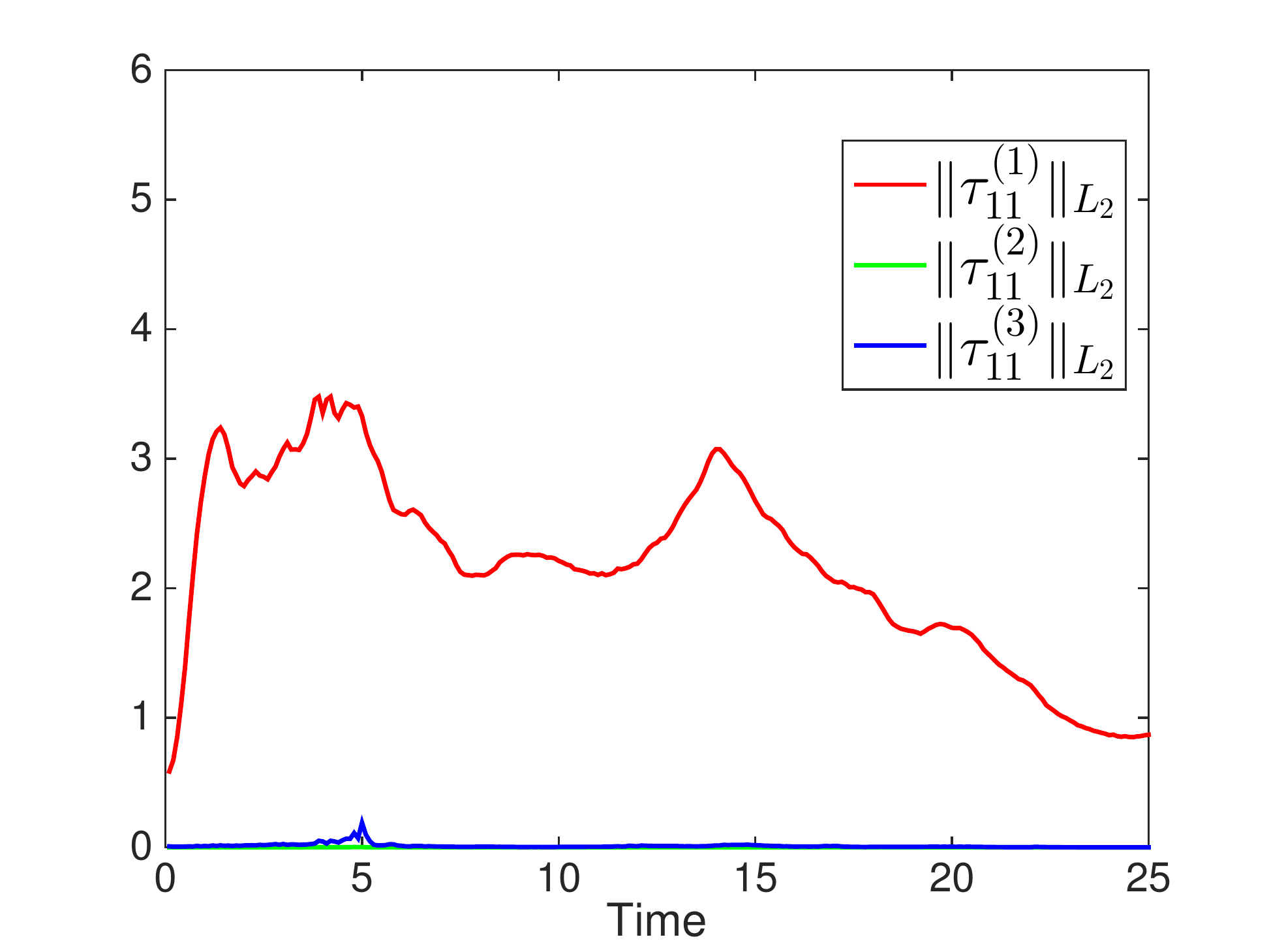}}
\end{subfigure}
\begin{subfigure}[]{
      \includegraphics[width=0.56\textwidth]{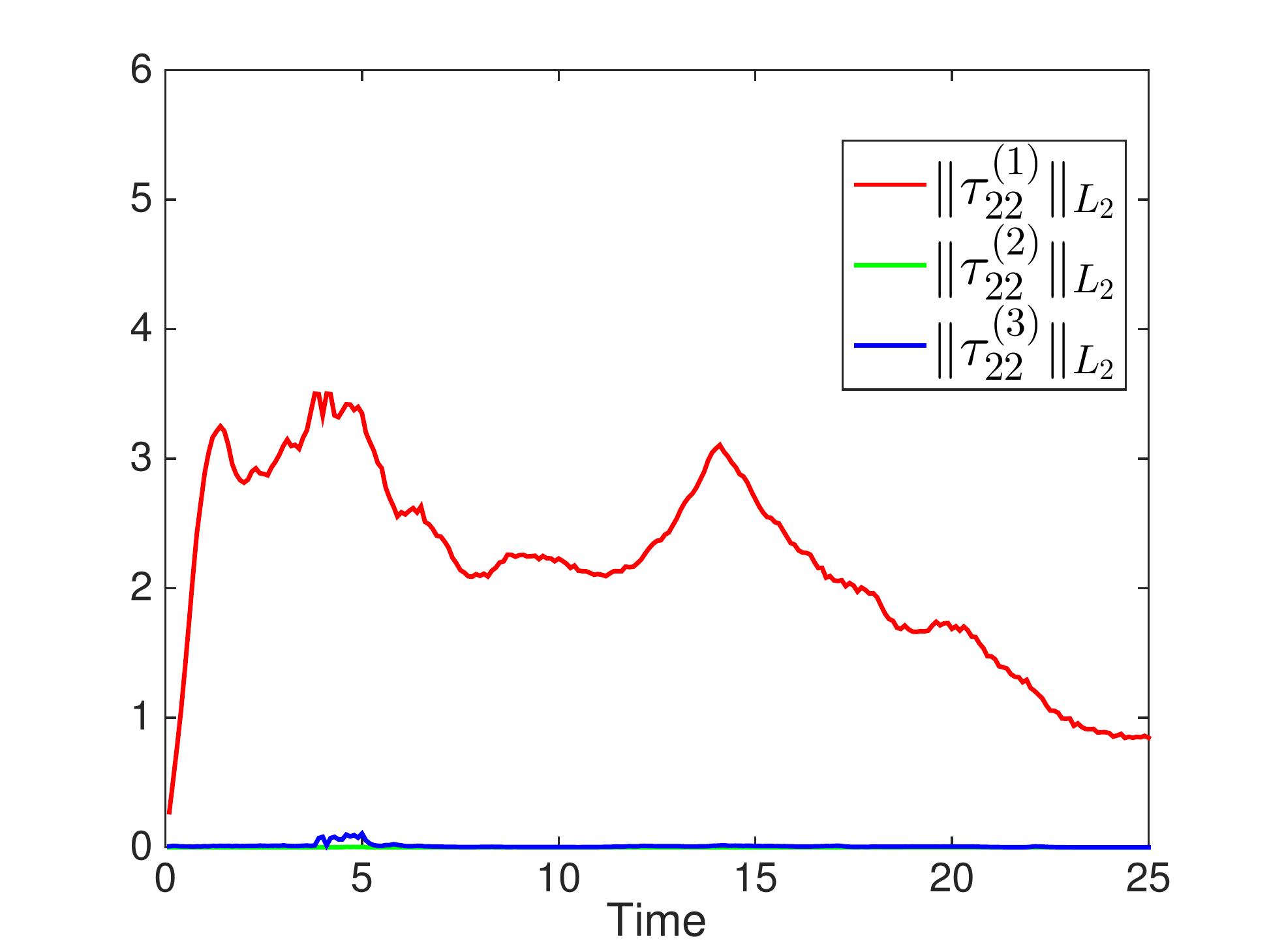}}
\end{subfigure} 
\begin{subfigure}[]{
       \includegraphics[width=0.56\textwidth]{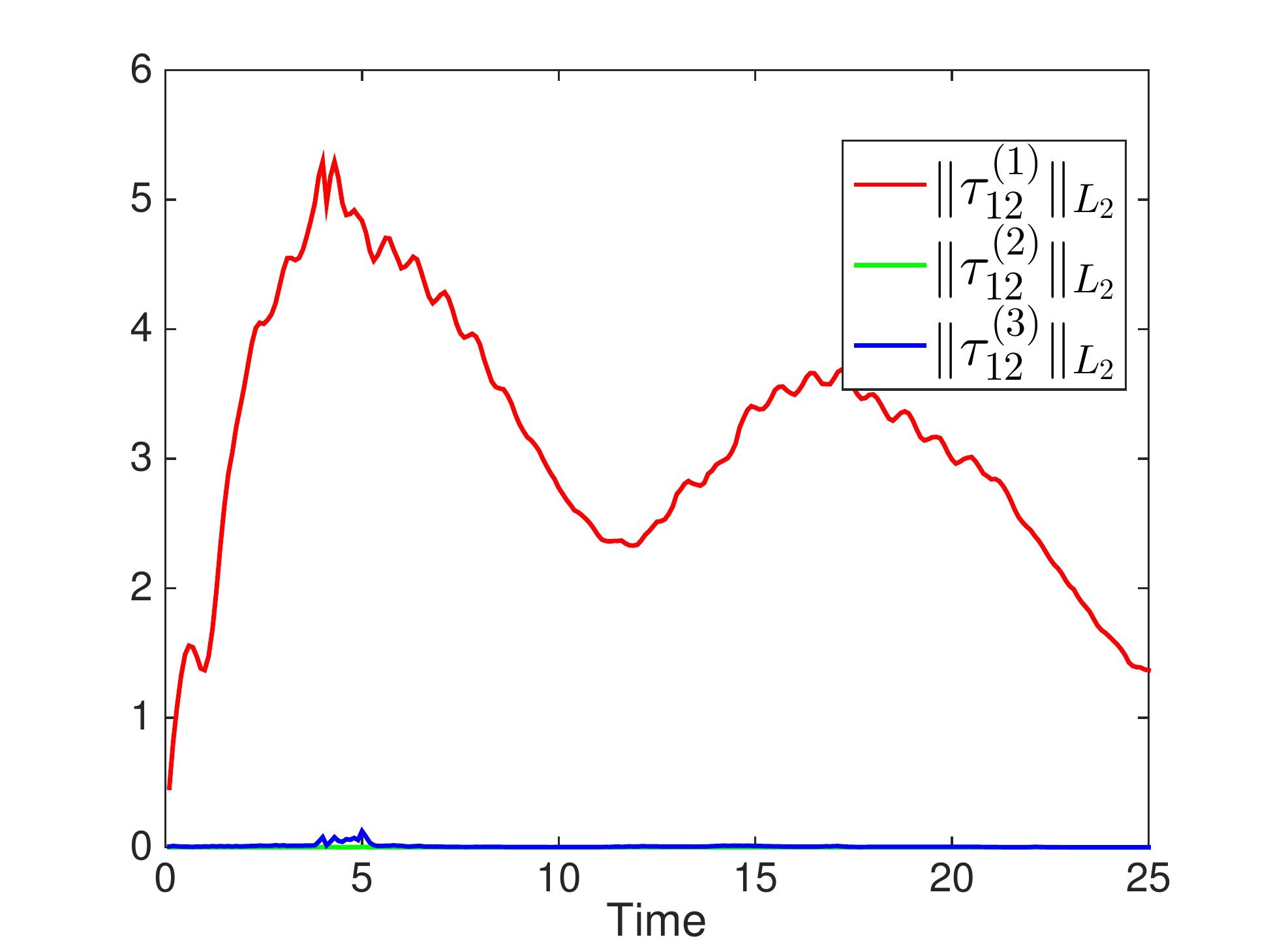}}
\end{subfigure}
\caption{$L^2$ norm of the different components of $\tau ^{(1)}$, $\tau ^{(2)}$ and $\tau^{(3)}$ as a function of time. (a) First diagonal component. (b) Second diagonal component. (c) Off diagonal component.}
\label{fig:l2norm_tau_contrib}
\end{figure}

If we now compare equations (\ref{eq:rel_favre_norm-b}) and (\ref{eq:theta_complete}), we can see that $\tau(u_i,u_j)$ can be written as the sum of the following three contributions:  
\begin{subequations}
\begin{align}
& \tau_{ij}^{(1)} = -\tilde S_{ij}, \label{eq:tau1}\\
& \tau_{ij}^{(2)} = -\frac{\tau(\rho,u_i)\partial_j \overline \rho + 
 \tau(\rho,u_j)\partial_i \overline \rho}{\overline \rho^2} ,\label{eq:tau2}\\
&\tau_{ij}^{(3)} =  \frac{\partial_j \tau(\rho,u_i)+\partial_i \tau(\rho, u_j)}{\overline \rho}, \label{eq:tau3} 
\end{align}
\label{eq:tau_contrib}
\end{subequations}
multiplied by $\nu_{\rm sgs}$. In figure \ref{fig:fr_norm_tau_contrib}, the Frobenius norm of the different terms (\ref{eq:tau_contrib}) is represented as a function of time. We notice that the contribution $\tau^{(1)}$ is much more important than the other two. A very similar trend is present in the $L_2$ norms of the different components of $\tau^{(1)}$, $\tau^{(2)}$ and $\tau^{(3)}, $ see figure \ref{fig:l2norm_tau_contrib}.

\begin{figure}[]
\centering
\begin{subfigure}[]{
      \includegraphics[width=0.56\textwidth]{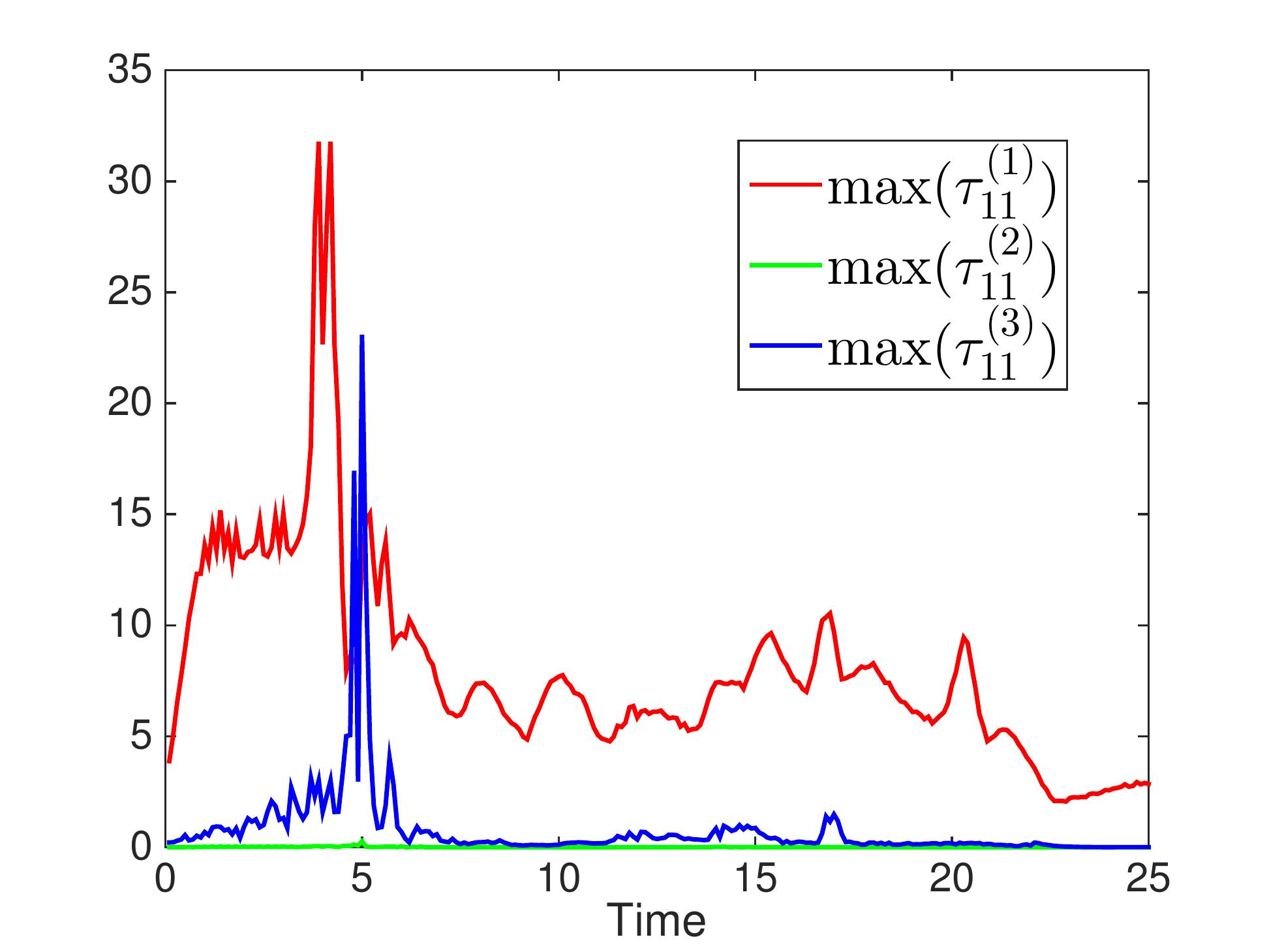}}
\end{subfigure}
\begin{subfigure}[]{
      \includegraphics[width=0.56\textwidth]{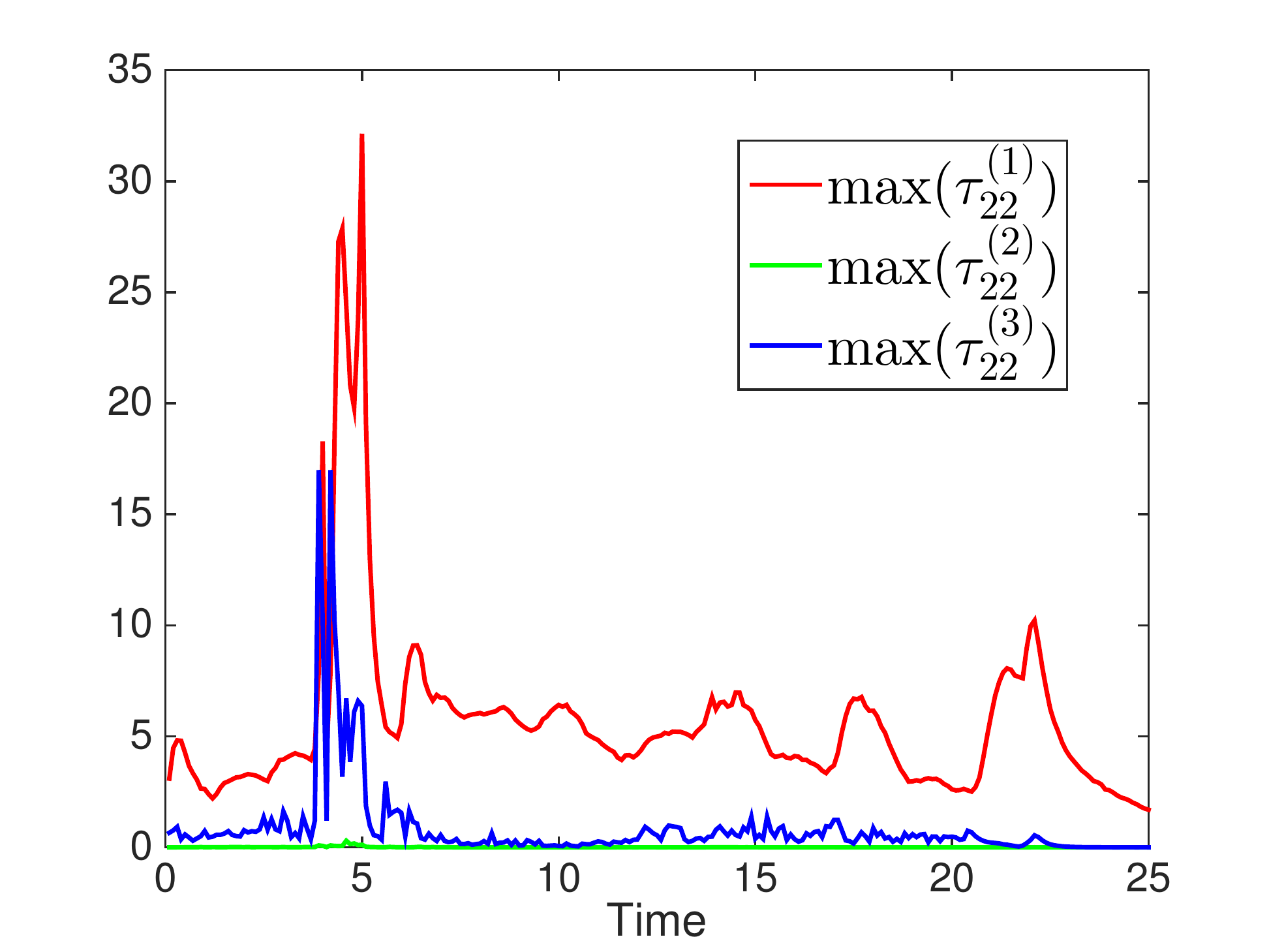}}
\end{subfigure} 
\begin{subfigure}[]{
       \includegraphics[width=0.56\textwidth]{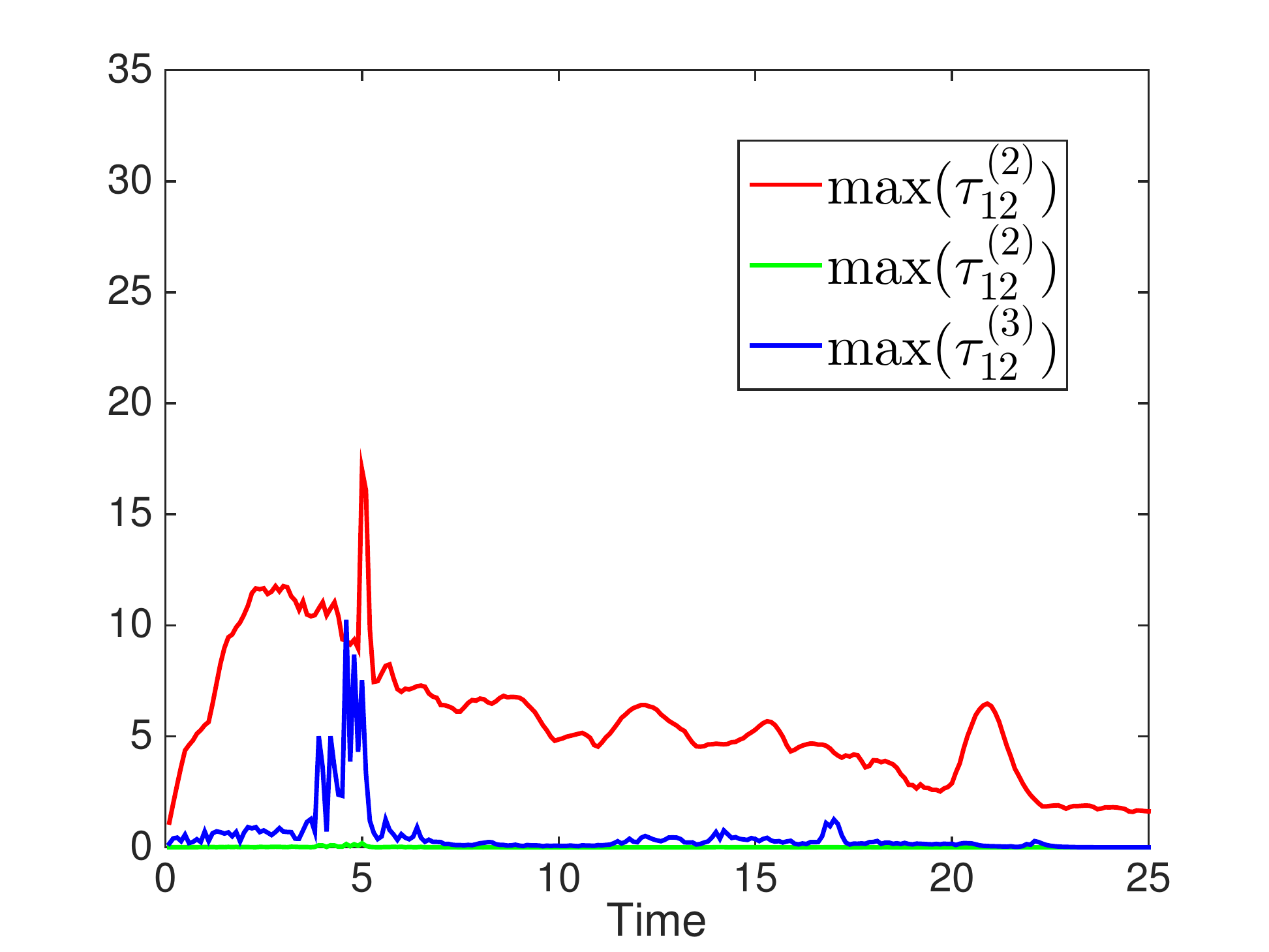}}
\end{subfigure}
\caption{Maximum value over the domain $\Omega$ of $\tau^{(1)}$, $\tau^{(2)}$ and $\tau^{(3)}$ as a function of time. (a) First diagonal component. (b) Second diagonal component. (c) Off-diagonal component.}
\label{fig:max_tau}
\end{figure}

\begin{figure}[]
\centering
\begin{subfigure}[]{
      \includegraphics[width=0.56\textwidth]{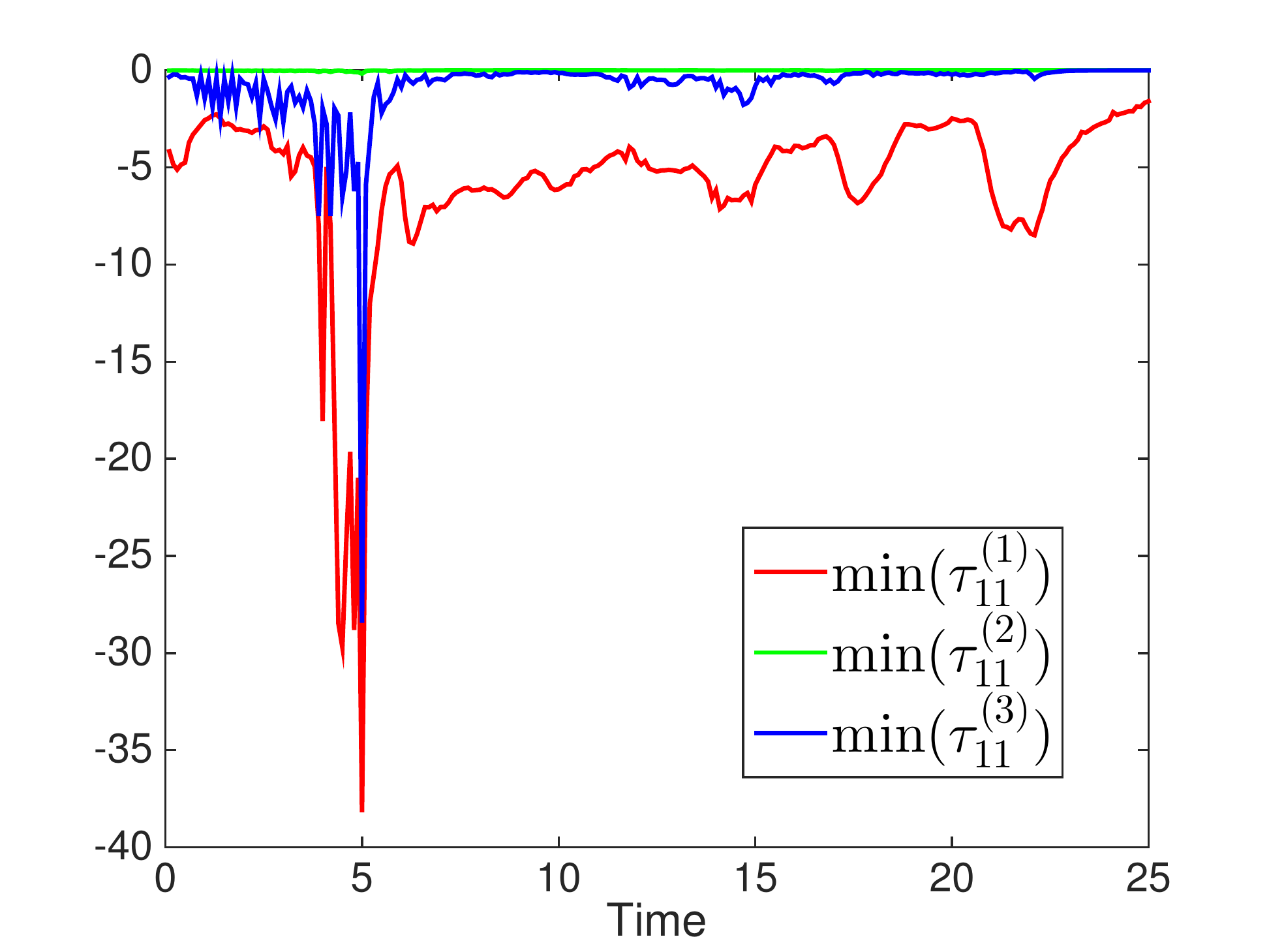}}
\end{subfigure}
\begin{subfigure}[]{
      \includegraphics[width=0.56\textwidth]{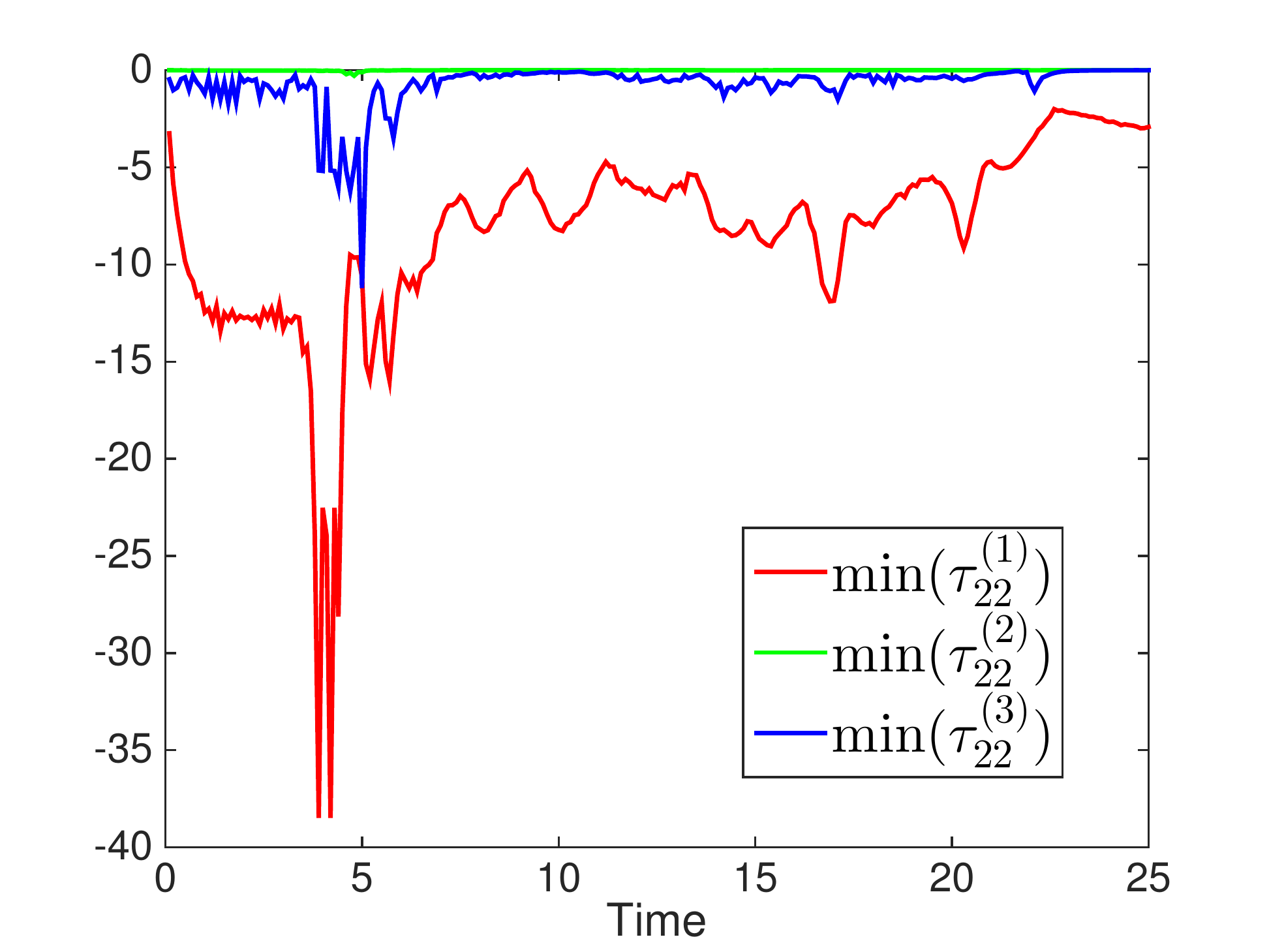}}
\end{subfigure} 
\begin{subfigure}[]{
       \includegraphics[width=0.56\textwidth]{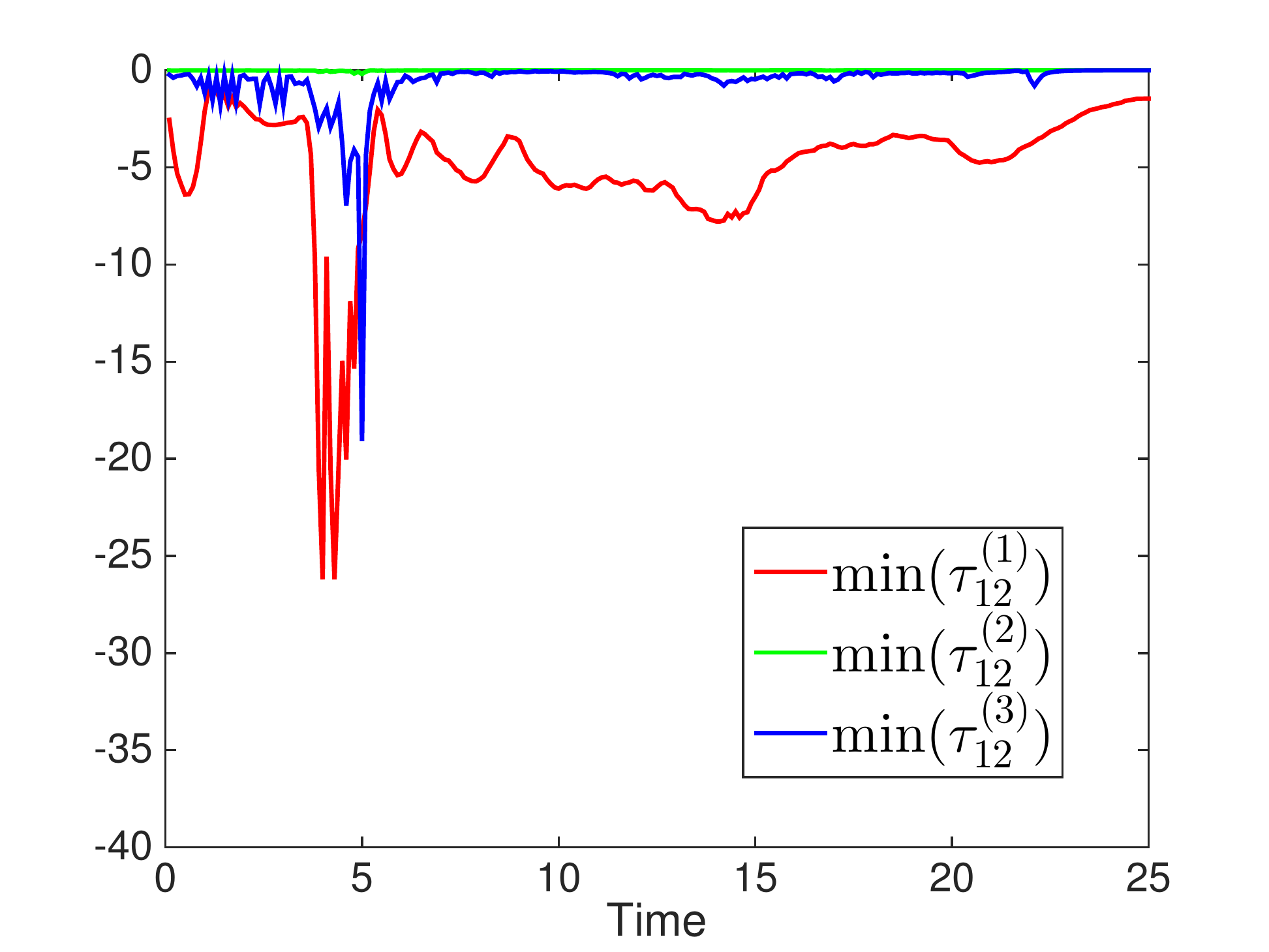}}
\end{subfigure}
\caption{Minimum value over the domain $\Omega$ of $\tau^{(1)}$, $\tau^{(2)}$ and $\tau^{(3)}$ as a function of time. (a) First diagonal component. (b) Second diagonal component. (c) Off-diagonal component.}
\label{fig:min_tau}
\end{figure}

In figures \ref{fig:max_tau} and \ref{fig:min_tau}, respectively, we show the time evolution of the maximum and minimum values over the domain $\Omega$ of $\tau^{(1)}$, $\tau^{(2)}$ and $\tau^{(3)}$. With respect to the evaluation in the Frobenius norm (see figure \ref{fig:fr_norm_tau_contrib}), we observe a more important contribution of $\tau^{(3)}$. 

Concluding, if we consider equation (\ref{eq:theta_complete}), the terms which are not negligible are:
\begin{subequations}
\begin{align}
& -\nu_{\rm sgs}\tilde S_{ij}, \label{eq:non_neg1}\\
& \nu_{\rm sgs} \tau^{(3)}_{ij} = \nu_{\rm sgs} \frac{\partial_j \tau(\rho,u_i)+\partial_i \tau(\rho, u_j)}{\overline \rho}, \label{eq:non_neg2}\\
& c_{ij}=\frac{\tau(\rho, u_i, u_j)}{\frho} \label{eq:non_neg3}.
\end{align}
\end{subequations}
Notice that, in addition to the first term (\ref{eq:non_neg1}), which is the only one usually considered in the traditional approach with density weighting for filtering in the compressible flows context, on the basis of the \textit{a priori} tests, also the terms (\ref{eq:non_neg2}) and (\ref{eq:non_neg3}) have to be retained when strong density gradients are present. 
 
This is in contrast with the modeling hypotheses proposed in \cite{germano:2014}, which are recalled here:
\begin{equation}
\nu_{\rm \rho u} = \nu_{\rm sgs}, \quad \nu_{\rm \rho} \neq 2\nu_{\rm sgs}.
\label{eq:germano_hyp}
\end{equation}
These hypotheses have the consequence that the terms $\nu_{\rm sgs}\tau^{(3)}$ (see equation (\ref{eq:non_neg2})) and $c_{ij}$ (equation (\ref{eq:non_neg3})), which are both non negligible according to the \textit{a priori} tests, cancel each other. Notice also that the hypothesis (\ref{eq:germano_hyp}) lead to the fact that the two terms $\nu_{\rm sgs}\tau^{(2)}$ (equation (\ref{eq:tau2})) and $b_{ij}$ (equation (\ref{eq:apriori_contrib-2})), which are negligible according to the \textit{a priori} tests, are retained in the Germano formulation. 

\section{Alternative modeling hypothesis}
\label{alt_mod}
In the previous section we have verified that, in addition to $-\nu_{\rm sgs}\fS_{ij}$, there are other important terms in the expression for the subgrid scale Favre stress, when dealing with flows characterized by strong density variations. However, we have also verified that some of the modeling hypotheses in \cite{germano:2014} are not in good agreement with the previous results of the \textit{a priori} tests. Another limitation of the approach in \cite{germano:2014} is that a third order moment, which is difficult to model, is introduced in the expression for the subgrid scale Favre stress. 
As a consequence, we try to propose an alternative modeling hypothesis and to verify its validity by means of \textit{a priori} tests.

\begin{figure}
\centering
\includegraphics[width=0.7\textwidth]{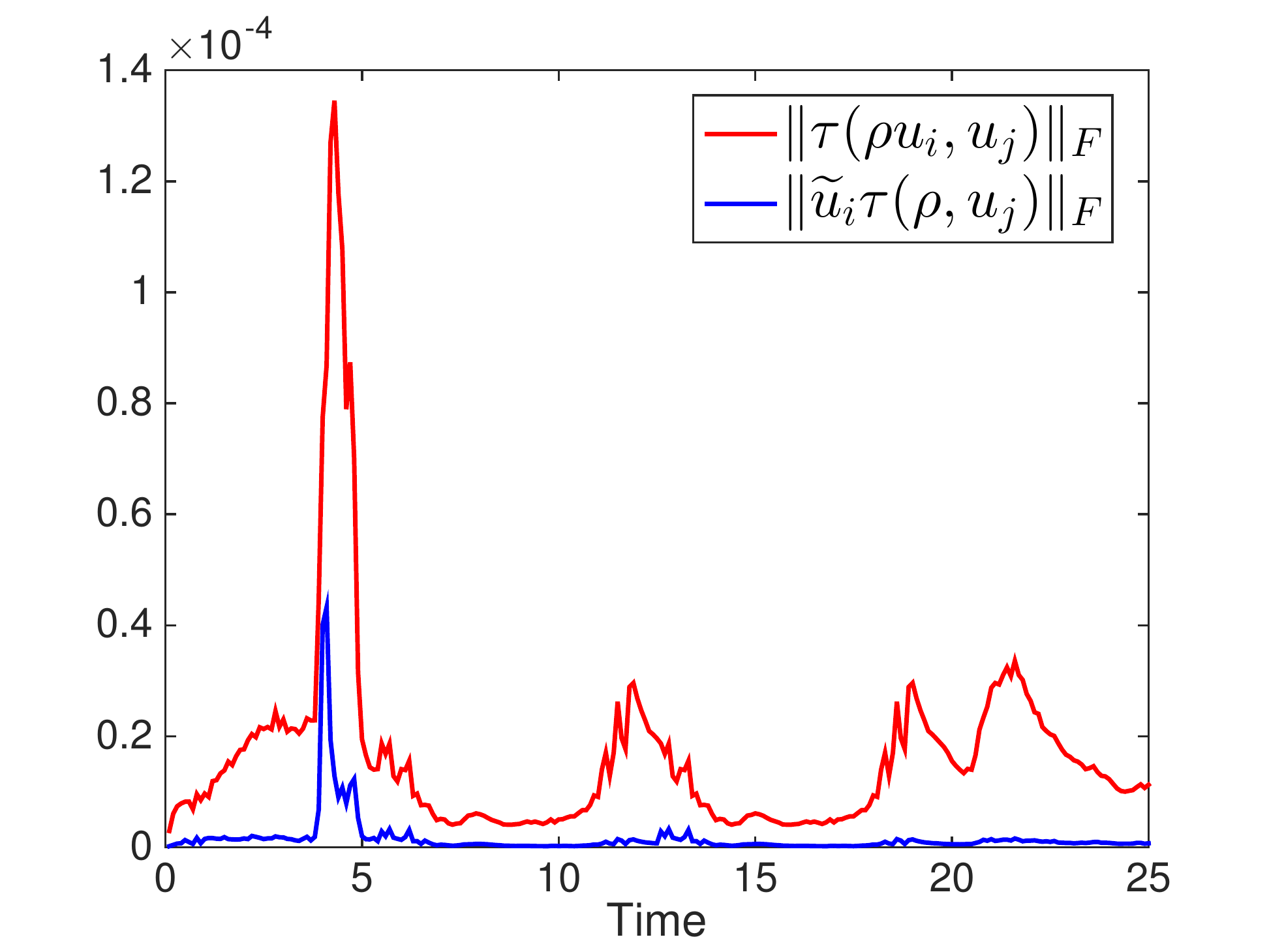}
\caption{Frobenius norm of $\tau(\rho u_i,u_j)$ and $\wt{u}_i \tau(\rho,u_j)$ as a function of time.}
\label{fig:newprop_fnorm}
\end{figure}

\begin{figure}[]
\centering
\begin{subfigure}[]{
      \includegraphics[width=0.48\textwidth]{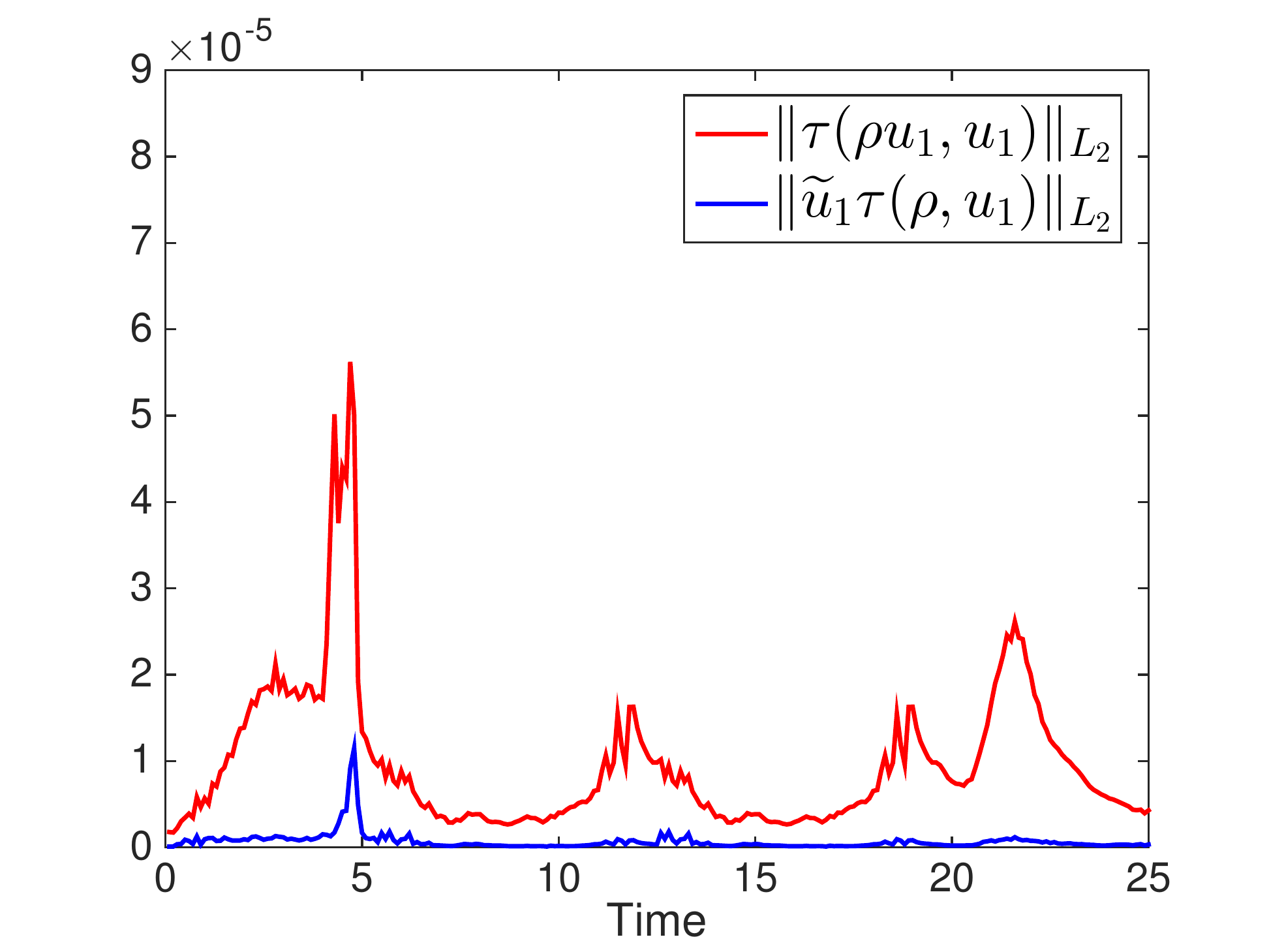}}
\end{subfigure}
\begin{subfigure}[]{
      \includegraphics[width=0.48\textwidth]{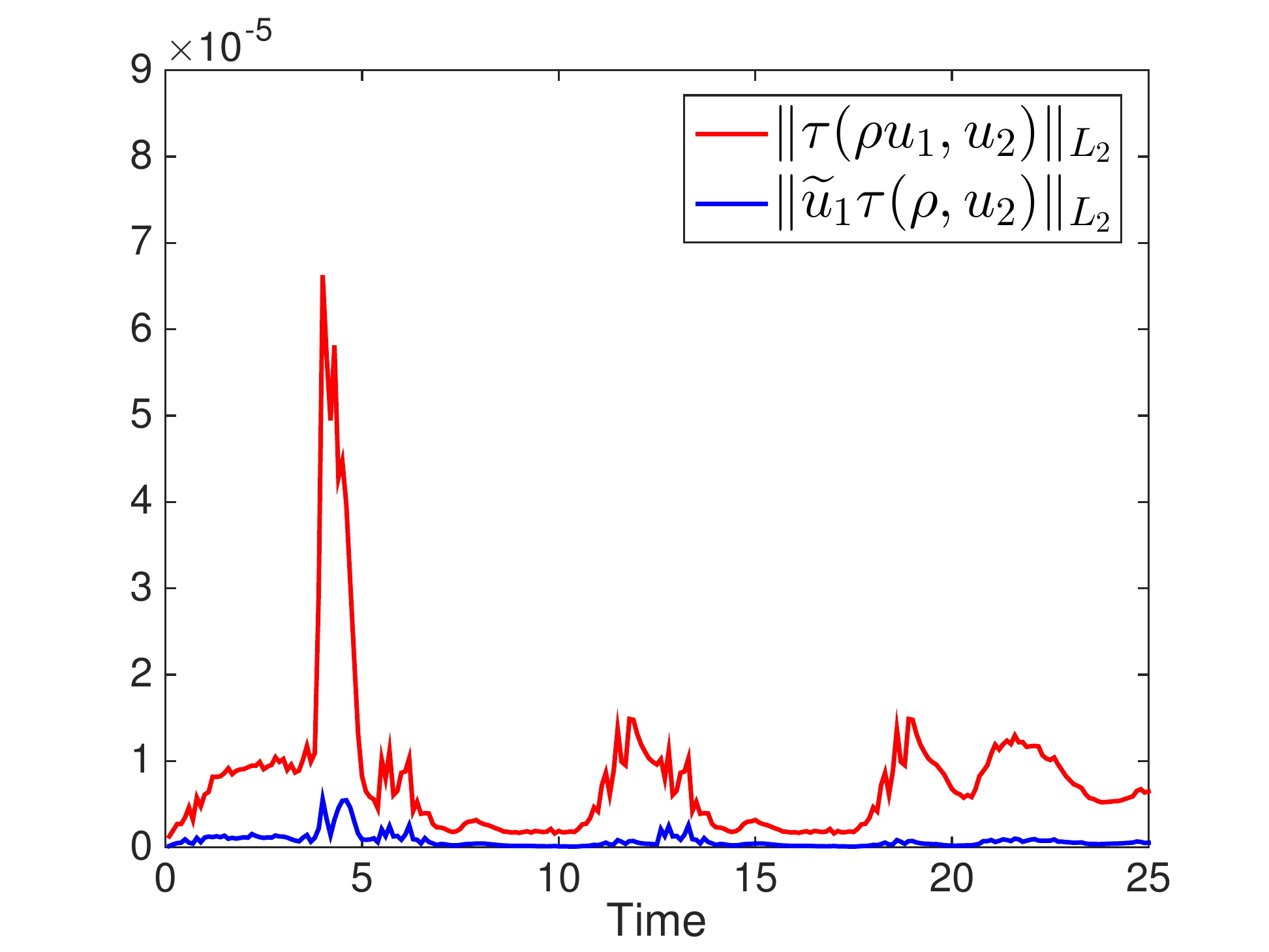}}
\end{subfigure} 
\begin{subfigure}[]{
       \includegraphics[width=0.48\textwidth]{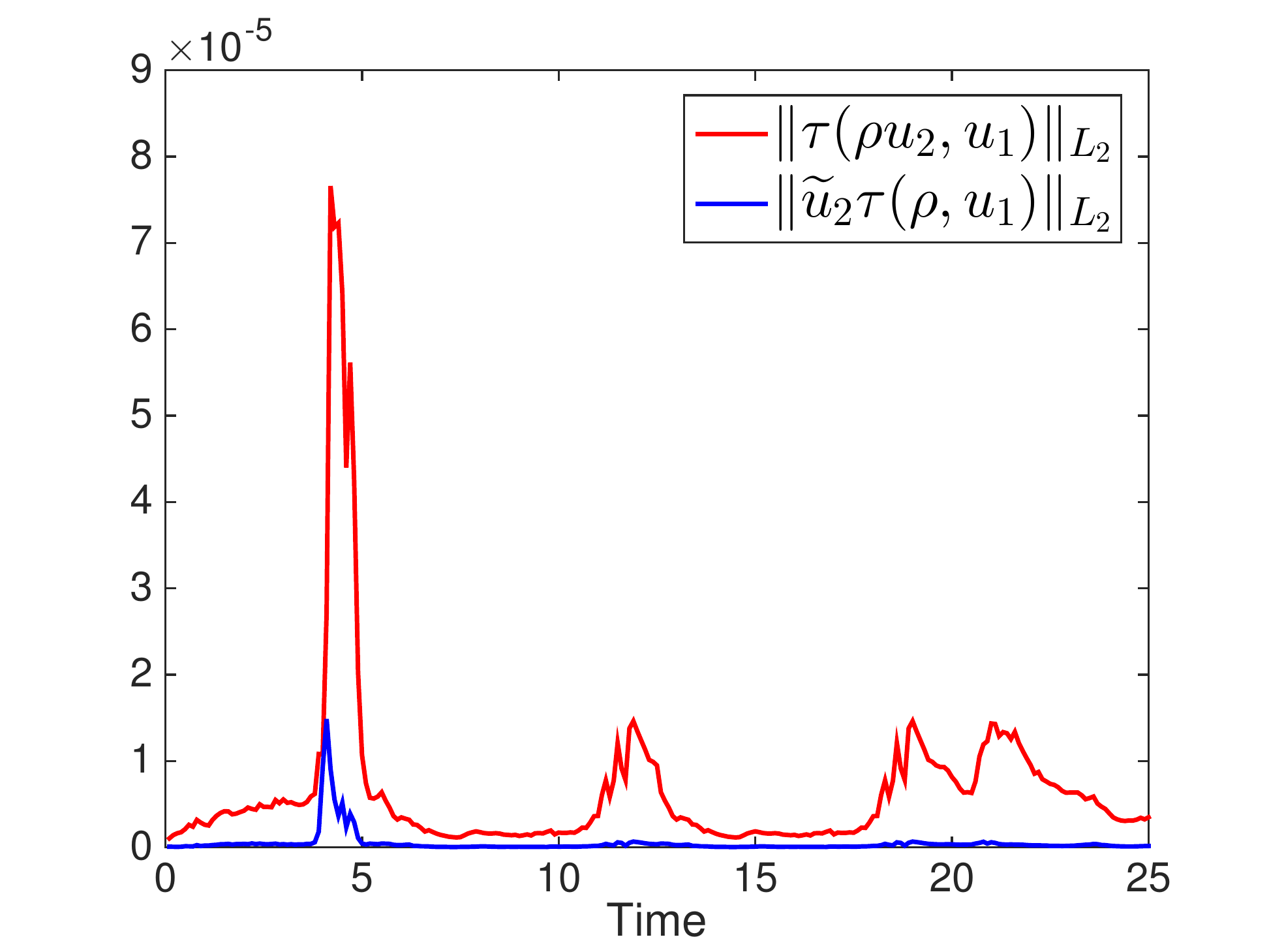}}
\end{subfigure}
\begin{subfigure}[]{
       \includegraphics[width=0.48\textwidth]{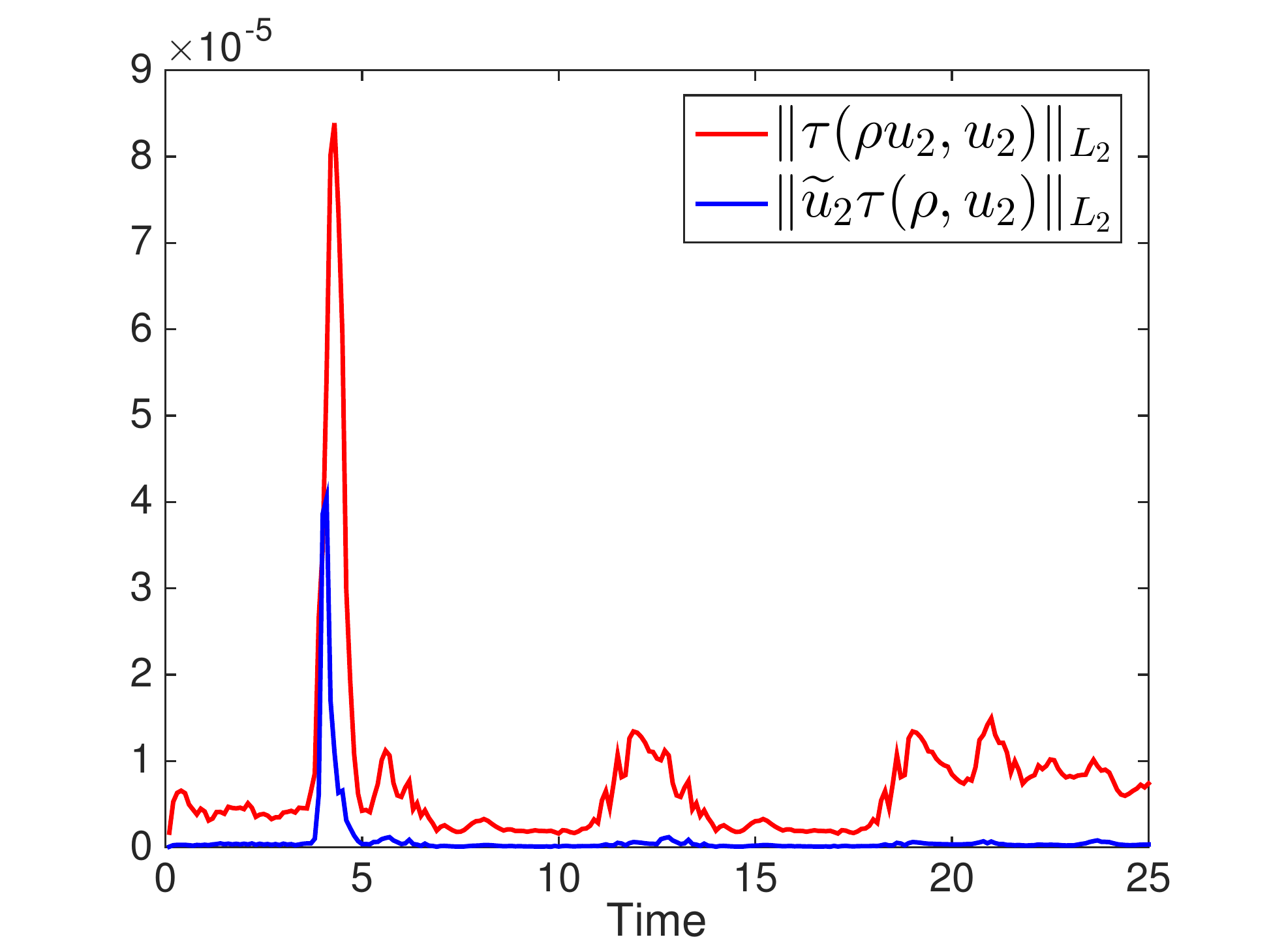}}
\end{subfigure}
\caption{$L^2$ norm of $\tau(\rho u_i,u_j)$ and $\wt{u}_i \tau(\rho,u_j)$ as a function of time. (a) Component $11$. (b) Component $12$. (c) Component $21$. (d) Component $22$.}
\label{fig:newprop_l2norm}
\end{figure}

\begin{figure}[]
\centering
\begin{subfigure}[]{
      \includegraphics[width=0.48\textwidth]{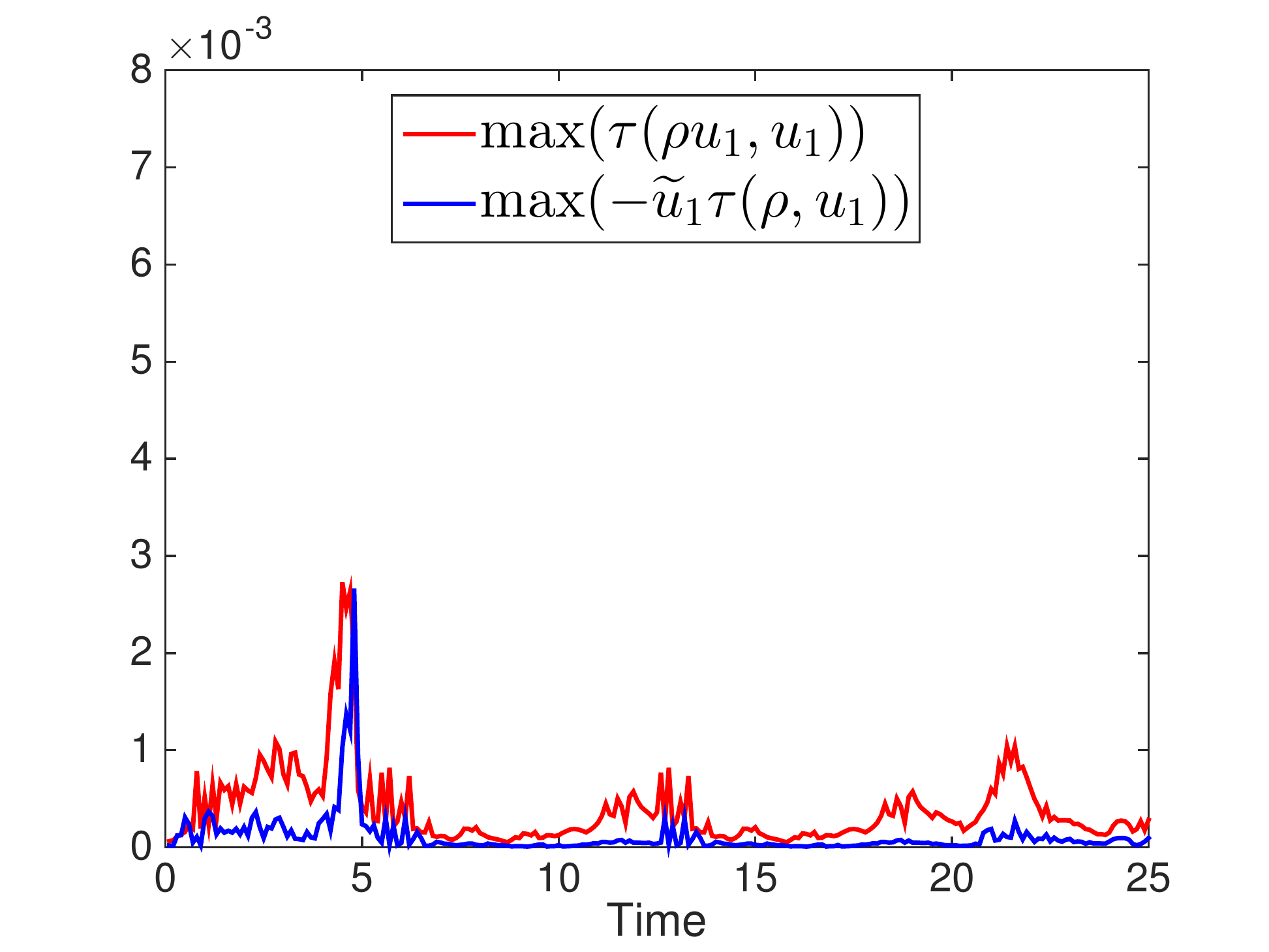}}
\end{subfigure}
\begin{subfigure}[]{
      \includegraphics[width=0.48\textwidth]{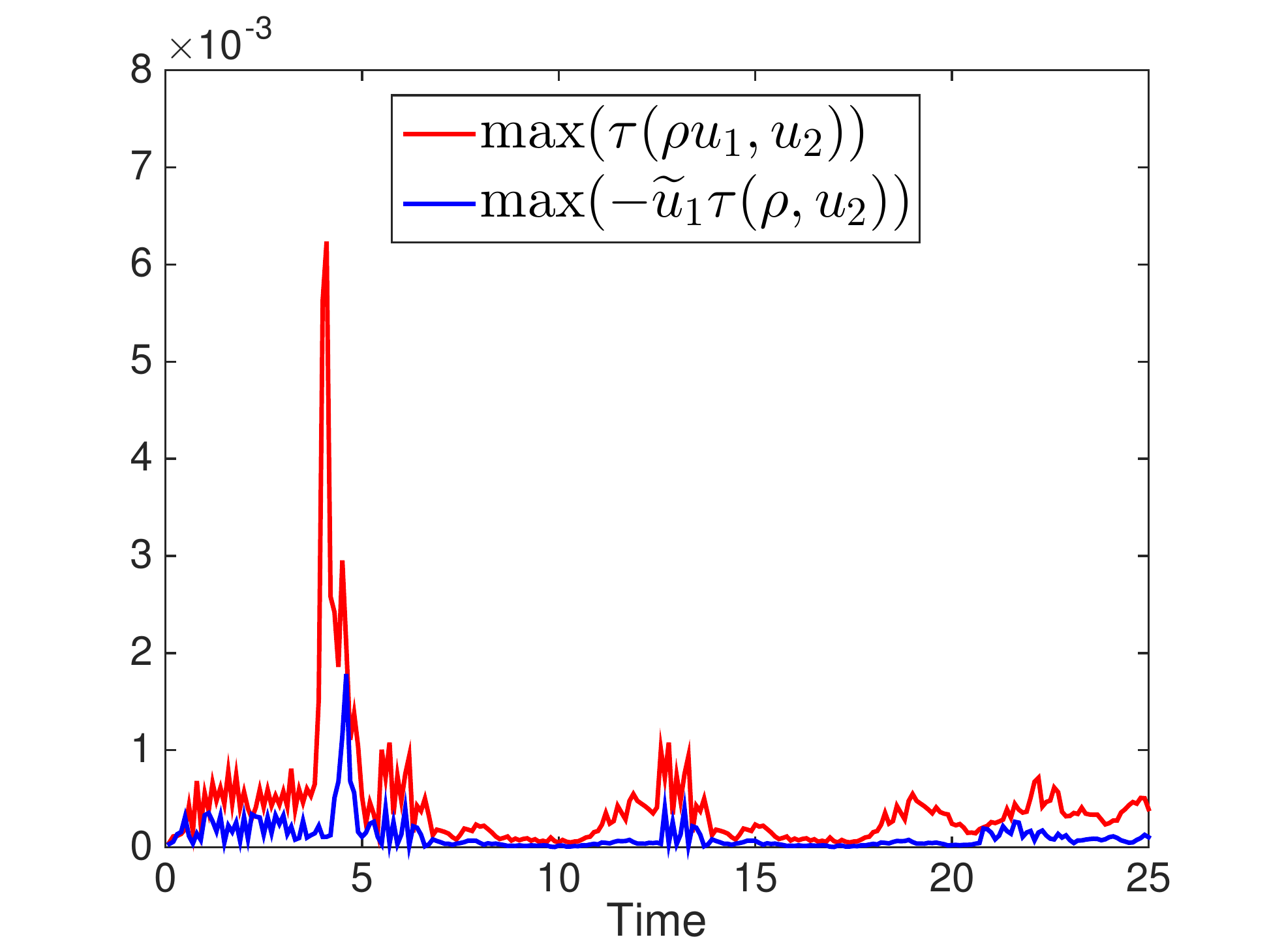}}
\end{subfigure} 
\begin{subfigure}[]{
       \includegraphics[width=0.48\textwidth]{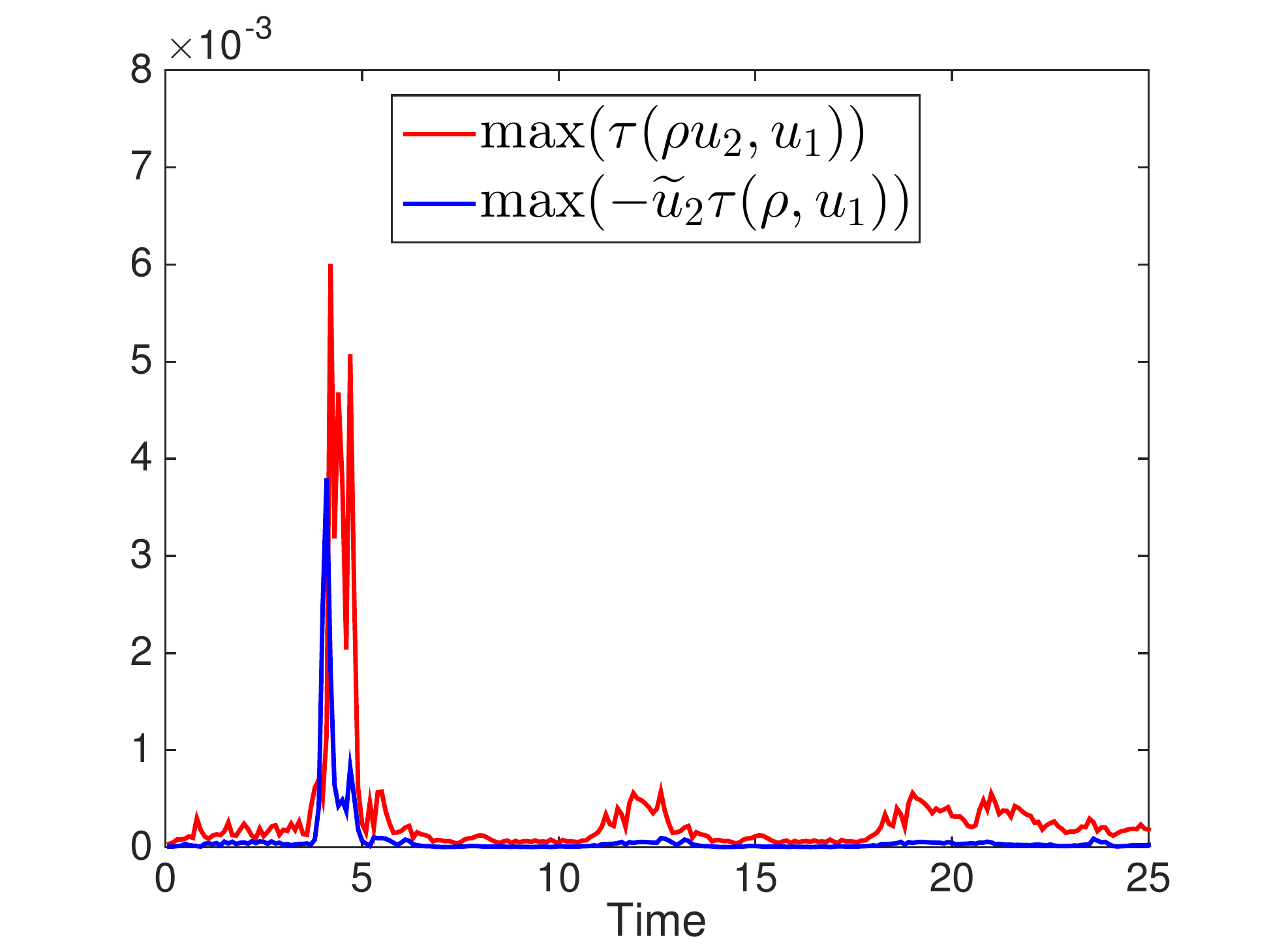}}
\end{subfigure}
\begin{subfigure}[]{
       \includegraphics[width=0.48\textwidth]{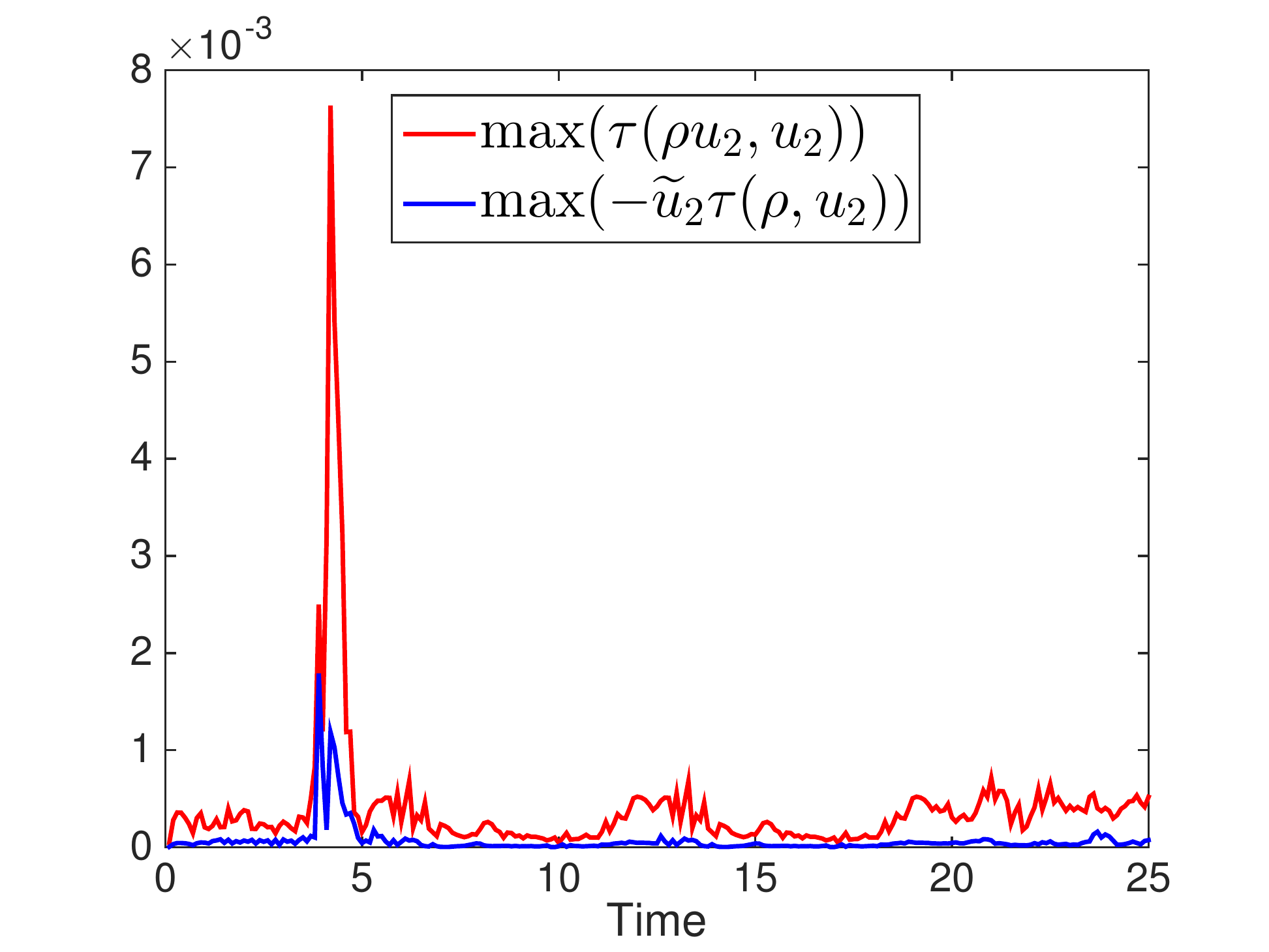}}
\end{subfigure}
\caption{Maximum value over the domain $\Omega$ of $\tau(\rho u_i,u_j)$ and $\wt{u}_i \tau(\rho,u_j)$ as a function of time. (a) Component $11$. (b) Component $12$. (c) Component $21$. (d) Component $22$.}
\label{fig:newprop_max}
\end{figure}

\begin{figure}[]
\centering
\begin{subfigure}[]{
      \includegraphics[width=0.48\textwidth]{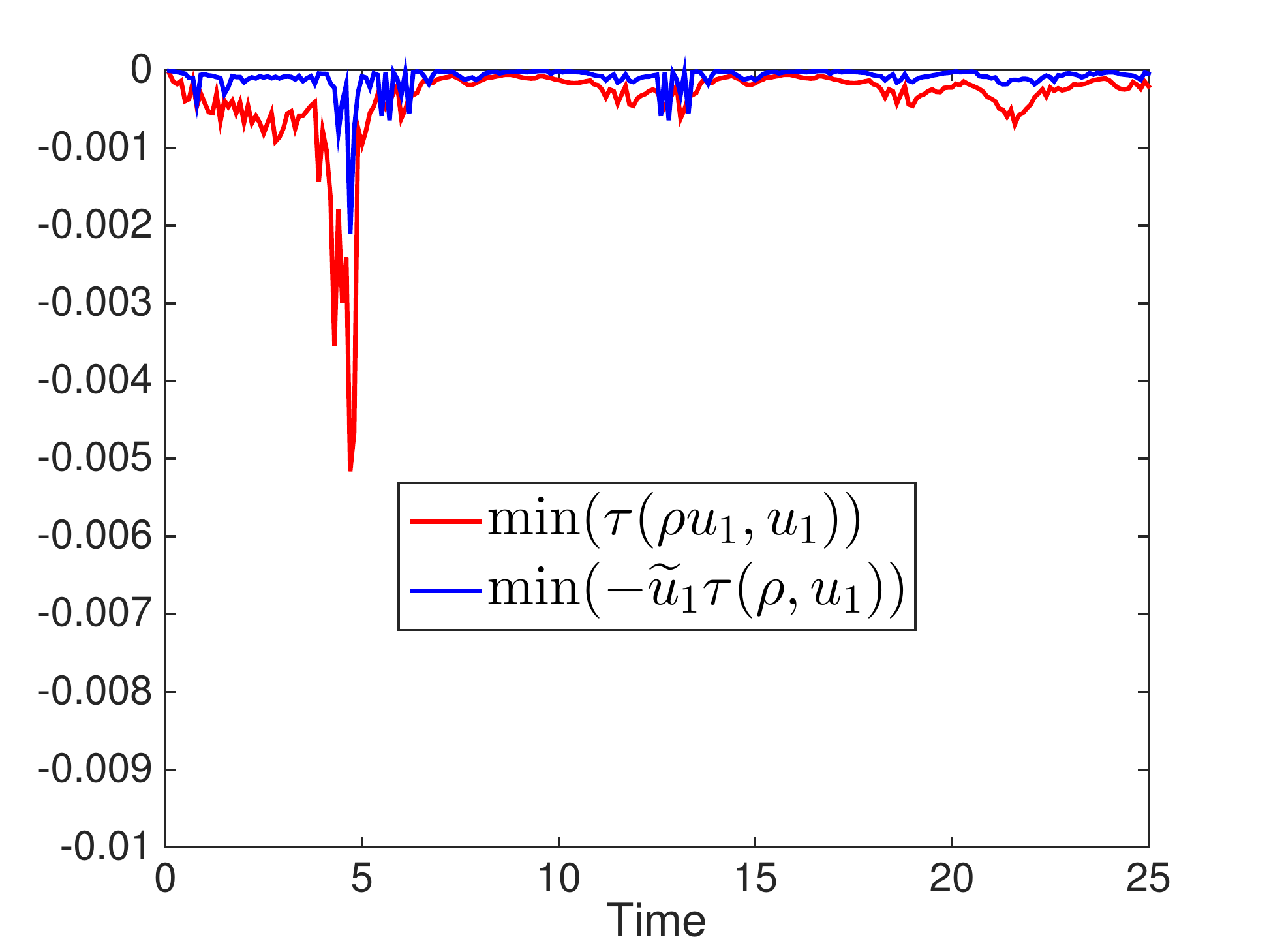}}
\end{subfigure}
\begin{subfigure}[]{
      \includegraphics[width=0.48\textwidth]{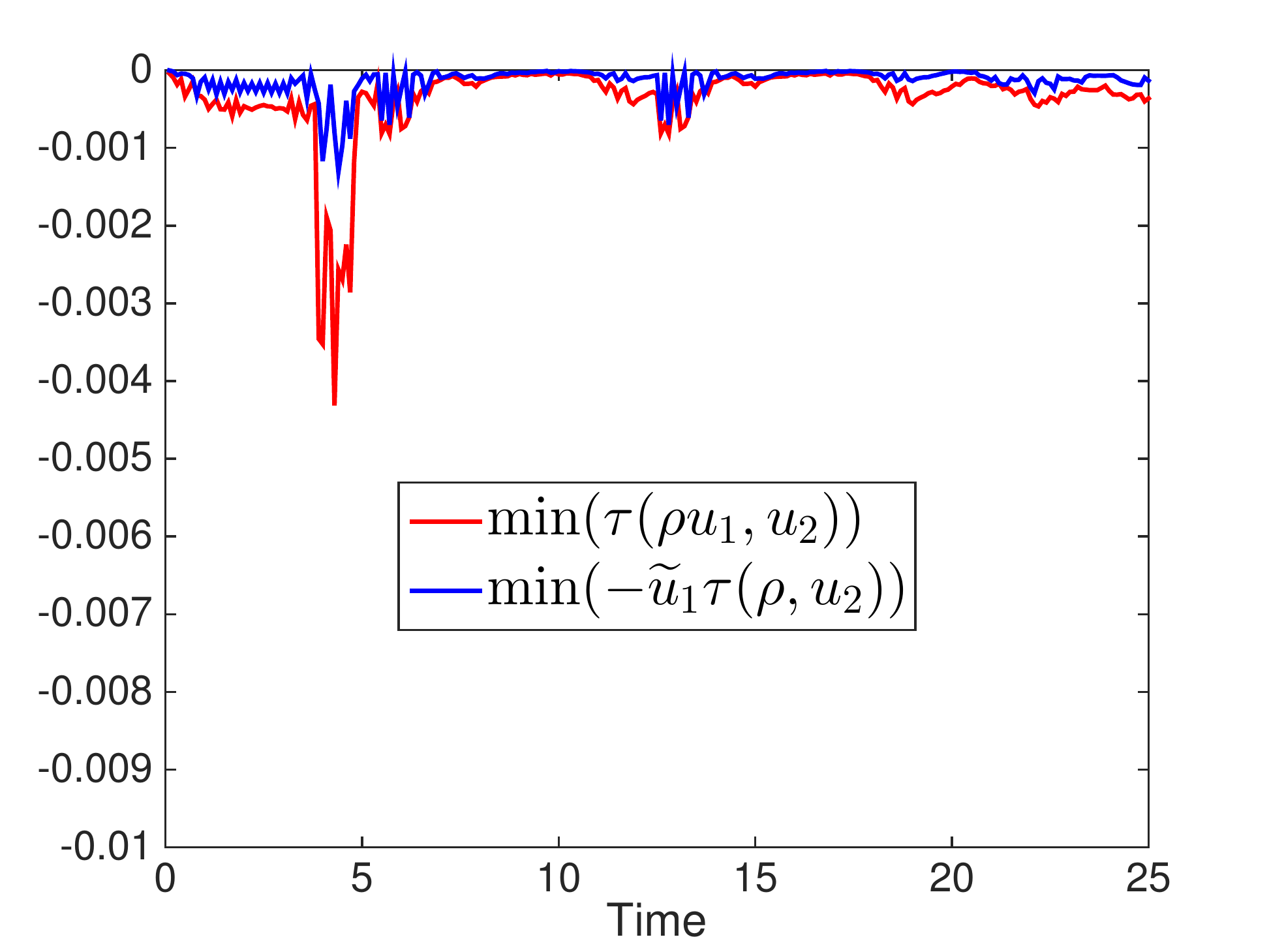}}
\end{subfigure} 
\begin{subfigure}[]{
       \includegraphics[width=0.48\textwidth]{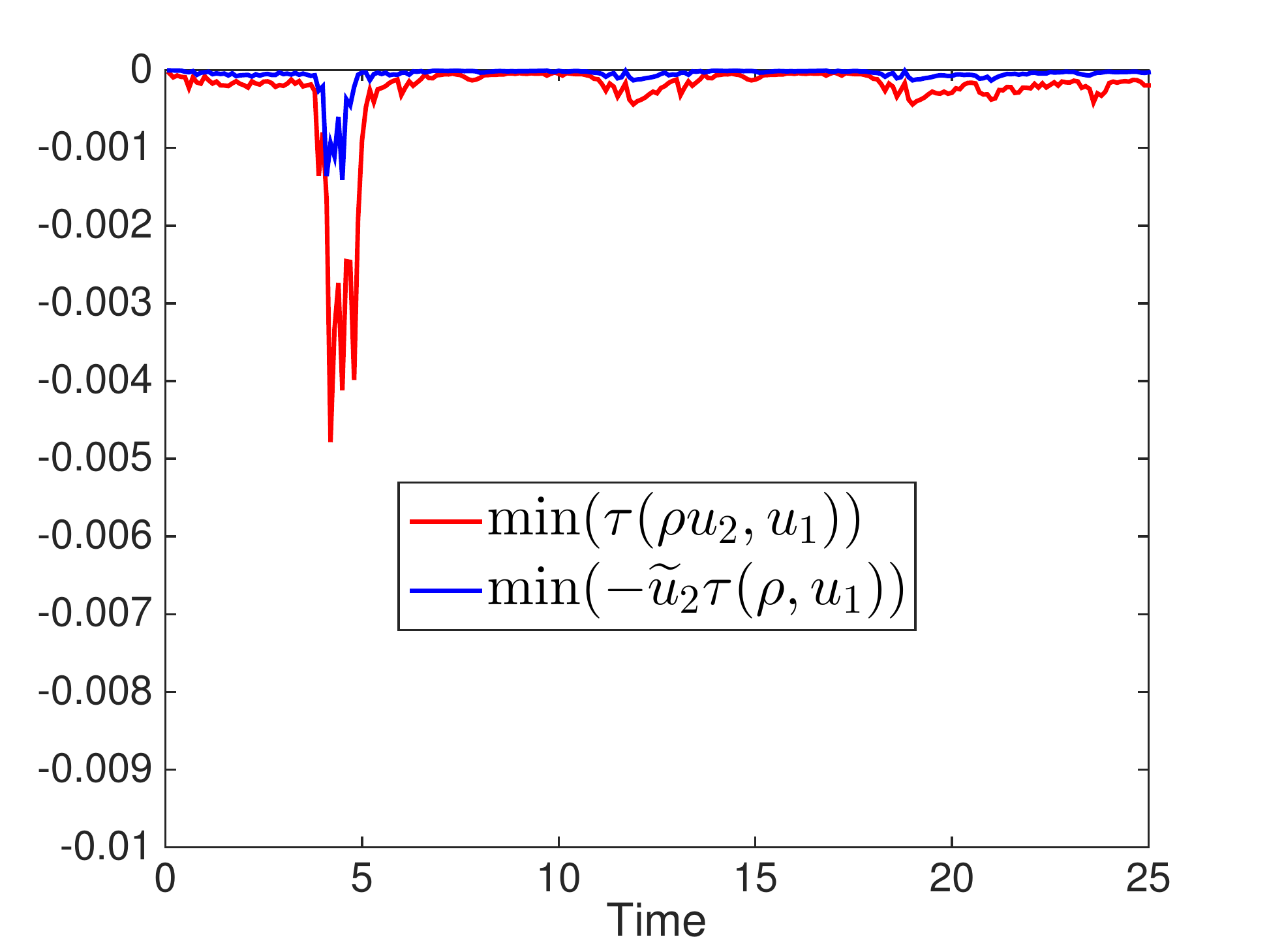}}
\end{subfigure}
\begin{subfigure}[]{
       \includegraphics[width=0.48\textwidth]{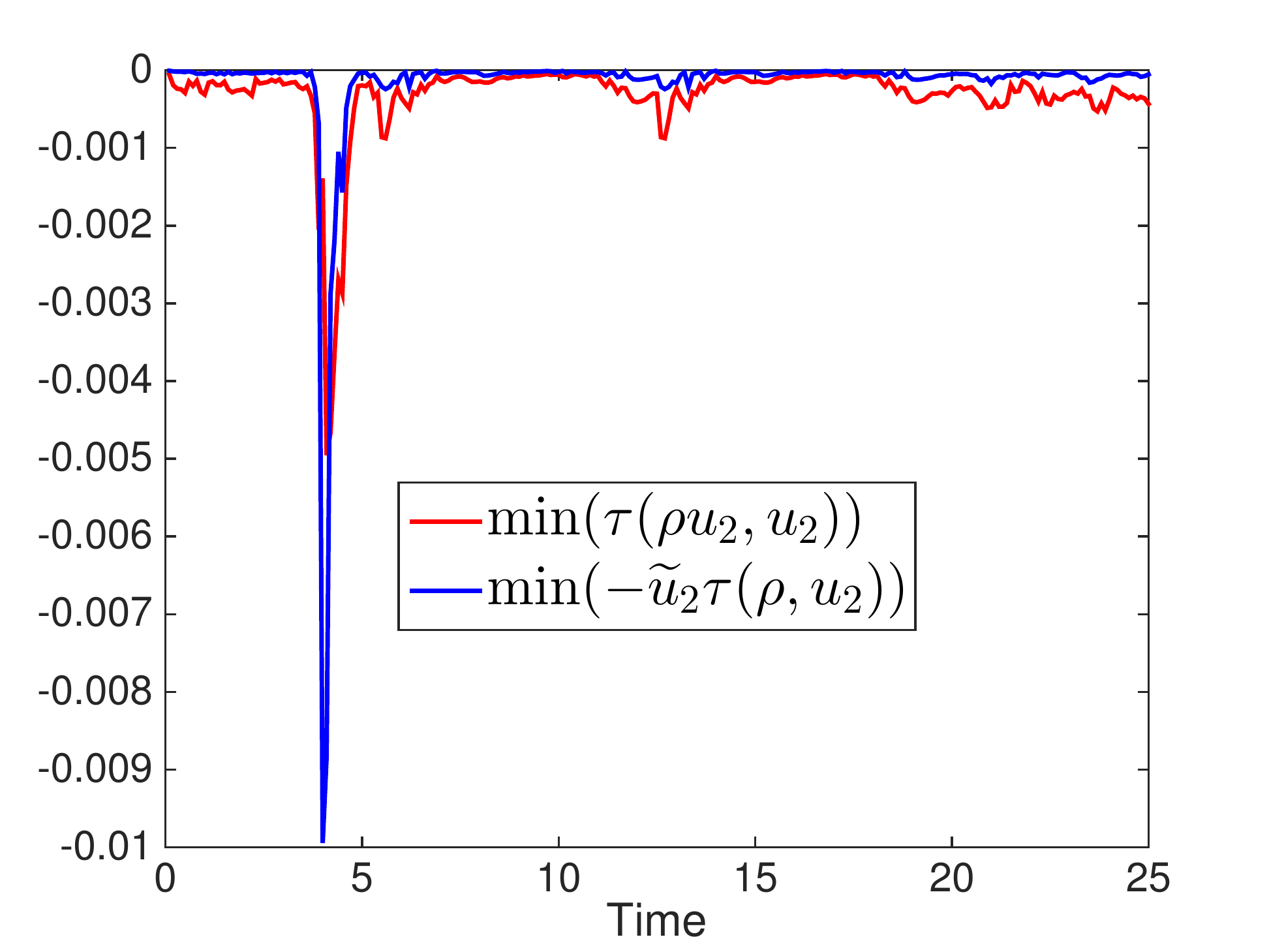}}
\end{subfigure}
\caption{Minimum value over the domain $\Omega$ of $\tau(\rho u_i,u_j)$ and $\wt{u}_i \tau(\rho,u_j)$ as a function of time. (a) Component $11$. (b) Component $12$. (c) Component $21$. (d) Component $22$.}
\label{fig:newprop_min}
\end{figure}

Using the definition \eqref{favreav} of Favre average  and substituting  it in equation (\ref{eq:frhoui}), we rewrite equation (\ref{eq:theta_base}) as:
\begin{eqnarray}
\overline {\rho} \theta(u_i,u_j)  &=&  \overline{\rho u_i u_j}- \overline {\rho} \wt{ u_i} \wt{ u_j}  \label{eq:rhothetauiuj}\\
&=&\frac{1}{2}\left[\overline{\rho u_i u_j}- \overline{\rho u_i}  \overline{u}_j  
+ \overline{\rho u_i u_j}- \overline{\rho u_j} \ \overline{u}_i
 \right. \nonumber \\
&+& \left. \ \ \ \ 
\overline{\rho u_i}  \overline{u}_j -
\overline {\rho} \wt u_i \wt u_j +
\overline{\rho u_j} \ \overline {u}_i -
\overline {\rho} \wt u_j \wt u_i\right]\nonumber \\
&  = & \frac{1}{2}\left[\tau(\rho u_i, u_j) + \tau(\rho u_j, u_i) - \overline \rho \wt u_i (\wt u_j -\bar u_j)
- \overline \rho \wt u_j (\wt u_i -\bar u_i)\right]\nonumber \\
&  = & \frac{1}{2}\left[\tau(\rho u_i, u_j) + \tau(\rho u_j, u_i) 
- \wt u_i \tau(\rho, u_j)- \wt u_j \tau(\rho, u_i)\right], \nonumber
\end{eqnarray}
where $\tau(\rho u_i, u_j)=\overline{\rho u_i u_j}- \overline{\rho u_i} \ \overline u_j$ is the subgrid flux of $\rho u_i$ advected by $u_j$.
As done in section (\ref{sec:results}) for the three contributions (\ref{eq:apriori_contrib}), we evaluate the time evolution of the Frobenius (figure \ref{fig:newprop_fnorm}) and the $L^2$ norms (figure \ref{fig:newprop_l2norm}) together with the time evolution of the maximum (figure \ref{fig:newprop_max}) and minimum values (figure \ref{fig:newprop_min}) of the terms $\tau(\rho u_i,u_j)$ and $-\wt{u_i}\tau(\rho,u_j)$ appearing in equation (\ref{eq:rhothetauiuj}). 
As it can be seen from figures \ref{fig:newprop_fnorm} and \ref{fig:newprop_l2norm}, the contribution of $-\wt{u_i}\tau(\rho,u_j)$, even if not negligible, is smaller than that of $\tau(\rho u_i,u_j)$. Figures \ref{fig:newprop_max} and \ref{fig:newprop_min}, where the time evolution of the maximum and minimum values of $\tau(\rho u_i,u_j)$ and $-\wt{u_i}\tau(\rho,u_j)$ over the whole domain is represented, further suggest that the term $-\wt{u_i}\tau(\rho,u_j)$ should be retained, since it provides a contribution which is not completely negligible with respect to $\tau(\rho u_i,u_j)$.

We propose two different modeling approaches for the terms $\tau(\rho,u_i)$ and $\tau(\rho u_i,u_j)$. The first approach is of eddy viscosity type, while the second one extends the similarity scale hypothesis, firstly proposed in \cite{bardina:1983} and successively extended to compressible flows in \cite{vreman:1995}, to compressible variable density flows. The two approaches are described in the following, together with the results of additional \textit{a priori} tests performed to verify the validity of these new hypotheses. 

\subsection{Eddy viscosity approach}
\begin{figure}
\centering
\includegraphics[width=0.7\textwidth]{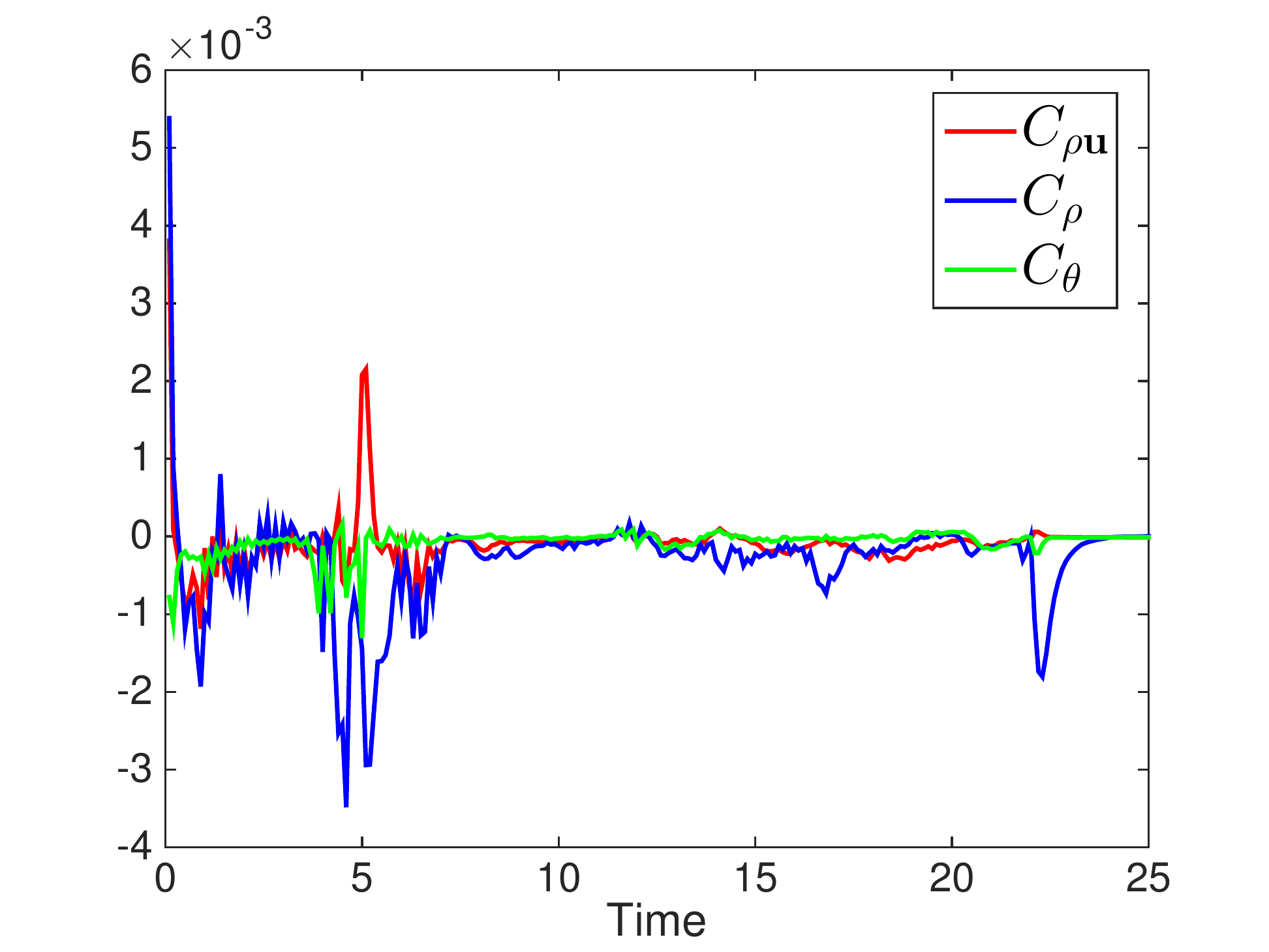}
\caption{Correlations $C_{\rho \bu}$, $C_\rho$ and $C_\theta$, corresponding to equations (\ref{eq:corr_mom}), (\ref{eq:corr}) and (\ref{eq:correlation_theta}), as a function of time.}
\label{fig:correlations}
\end{figure}

Considering an eddy viscosity approach, the two terms $\tau(\rho u_i, u_j)$  and $\tau(\rho,u_i)$ are modeled as:
\begin{subequations}
\begin{align}
&\tau(\rho u_i, u_j) =
-\nu_{\rm 1} \partial_j\overline{\rho u_i}  = -\nu_{\rm 1} \partial_j (\overline \rho \wt u_i), \label{eq:nu1}\\
& \tau(\rho, u_i) = -\nu_{\rm \rho}\partial_i\overline{\rho}.\label{eq:nurho}
\end{align}
\label{eq:eddy_viscosity_approach}
\end{subequations}
As a first \textit{a priori} test of this modelling assumption, we evaluate the correlations between  
 $\tau(\rho u_i, u_j)$ and $\de_j(\overline{\rho u_i})$ (see equation (\ref{eq:nu1})) and between $\tau(\rho,u_i)$ and $\de_i \frho$ (see equation (\ref{eq:nurho})), given respectively by:
\begin{subequations}
\begin{align}
& C_{\rho \bu} = \frac{\frac{1}{|\Omega|} \int_{\Omega} \tau(\rho u_i,u_j) \de_j \overline{\rho u_i} d\bx}{\sqrt{\frac{1}{|\Omega|^2} \int_{\Omega}|\tau(\rho u_i,u_j)|^2 d\bx \int_{\Omega}|\de_j(\overline{\rho u_i})|^2 d\bx}}, \label{eq:corr_mom}\\
& C_{\rho} = \frac{\frac{1}{|\Omega|} \int_{\Omega} \tau(\rho, u_i) \de_i \overline{\rho} d\bx}{\sqrt{\frac{1}{|\Omega|^2} \int_{\Omega}|\tau(\rho, u_i)|^2 d\bx \int_{\Omega}|\de_i\overline{\rho }|^2 d\bx}}. \label{eq:corr}
\end{align}
\label{eq:correlations}
\end{subequations}
In figure \ref{fig:correlations} these quantities are shown, together with the correlation between $\theta(u_i,u_j)$ and $\fS_{ij}$ given by the following equation:
\begin{equation}
C_{\theta} =  \frac{\frac{1}{|\Omega|} \int_{\Omega} \theta(u_i,u_j) \fS_{ij} d\bx}{\sqrt{\frac{1}{|\Omega|^2} \int_{\Omega}|\theta(u_i,u_j)|^2 d\bx\int_{\Omega}|\fS_{ij}|^2d\bx}}.
\label{eq:correlation_theta}
\end{equation}
The quantities $\theta(u_i,u_j)$ and $\fS_{ij}$ are those which are usually set proportional to each other in the conventional approach to turbulence modeling for compressible flows. 
Notice that the fact that mainly negative correlations arise is due to the fact that a minus sign is present on the right-hand side of equations (\ref{eq:tau_ij_compr}), (\ref{eq:nu1}) and (\ref{eq:nurho}).

The correlation between $\theta(u_i,u_j)$ and $\fS_{ij}$ is low in absolute value. Notice also that, even though the hypotheses (\ref{eq:nu1}) and (\ref{eq:nurho}) appear to improve the results  with respect to the traditional hypothesis (this is true in particular  for $C_\rho$), also $C_{\rho \bu}$ and $C_\rho$ remain low. 
On the other hand, low correlation values in  \textit{a priori} tests are somewhat typical for eddy viscosity models, as discussed for example in \cite{abba:2001} for the  case of a  turbulent channel flow benchmark. 

In order to try to obtain a simpler approach with respect to that of equations (\ref{eq:eddy_viscosity_approach}) and since we can notice that both the subgrid fluxes in equations (\ref{eq:eddy_viscosity_approach}) are advected by the velocity field $u_i$, we verify by means of additional \textit{a priori} tests if the simplification $\nu_{\rm 1} = \nu_{\rm \rho}$ can be introduced. Notice that the simplification $\nu_{\rm 1} = \nu_{\rm \rho}$ implicitly implies that we are considering scalar values for $\nu_{\rm 1}$ and $\nu_{\rm \rho}$. 
As a preliminar remark notice however that, if we assume $\nu_{\rm 1}=\nu_{\rm \rho}$, we go back to the conventional model $\overline{\rho} \theta(u_i,u_j) = -\overline{\rho}\nu_{\rm 1} \wt{\mathcal{S}}_{ij}$, where the only difference could be the introduction of an alternative expression for the eddy viscosity $\nu_{\rm 1}$ with respect to the conventional $\nu_{ \rm sgs} = C_S \Delta^2 |\wt{\mathcal{S}}|. $

In order to simply compare the two quantities $\nu_{1,ij}=\frac{\tau(\rho u_i,u_j)}{\de_j \overline{\rho u_i}}$ and $\nu_{\rho,i} = \frac{\tau(\rho,u_i)}{\de_i \overline{\rho}}$, we compute the two following expressions:
\begin{equation}
  \alpha_1=\frac{\Vert \tau(\rho u_i,u_j) \Vert_F}{\Vert \de_j \overline{\rho u_i}\Vert_F},  \ \ \ 
\alpha_{\rho}= \frac{\Vert \tau(\rho, u_i) \Vert_F}{\Vert \de_i \overline{\rho}\Vert_F}.
\label{eq:fnorm_nu1_nurho}
\end{equation}
 Notice that  we compute separately the Frobenius norms of the numerator and of the denominator in the expressions of $\nu_{1,ij}$ and $\nu_{\rho,i},$ in order not to have problems with integration points in which the modeled terms at the denominator become zero. This implies that $  \alpha_1 $ and $\alpha_{\rho}$  are just
 rough approximations of the size of  $\nu_{\rm 1}$ and $\nu_{\rm \rho}.$ 

\begin{figure}[]
\centering
\begin{subfigure}[\label{fig:nu1_nurho-a}]{
      \includegraphics[width=0.56\textwidth]{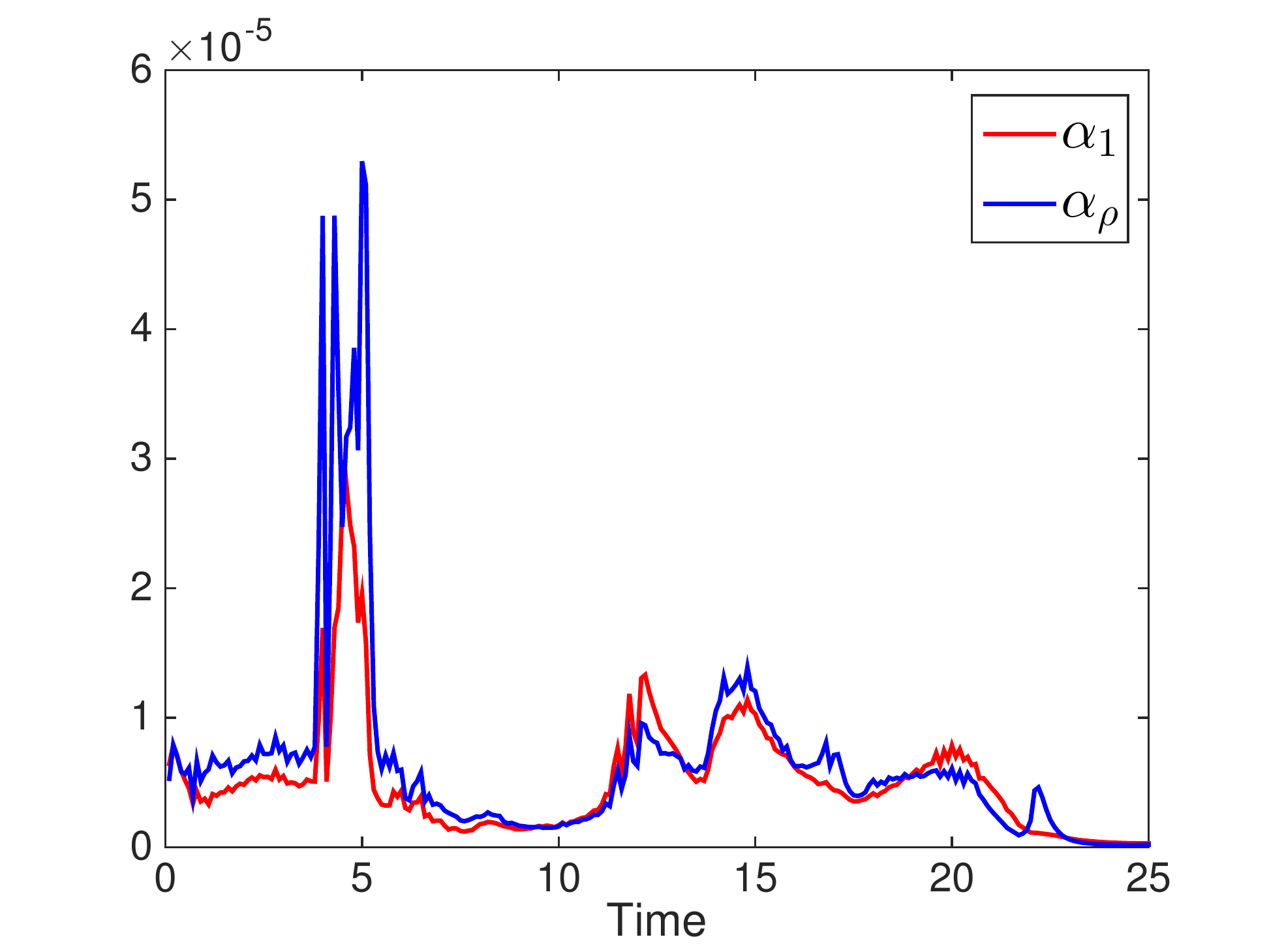}}
\end{subfigure}
\begin{subfigure}[\label{fig:nu1_nurho-b}]{
      \includegraphics[width=0.56\textwidth]{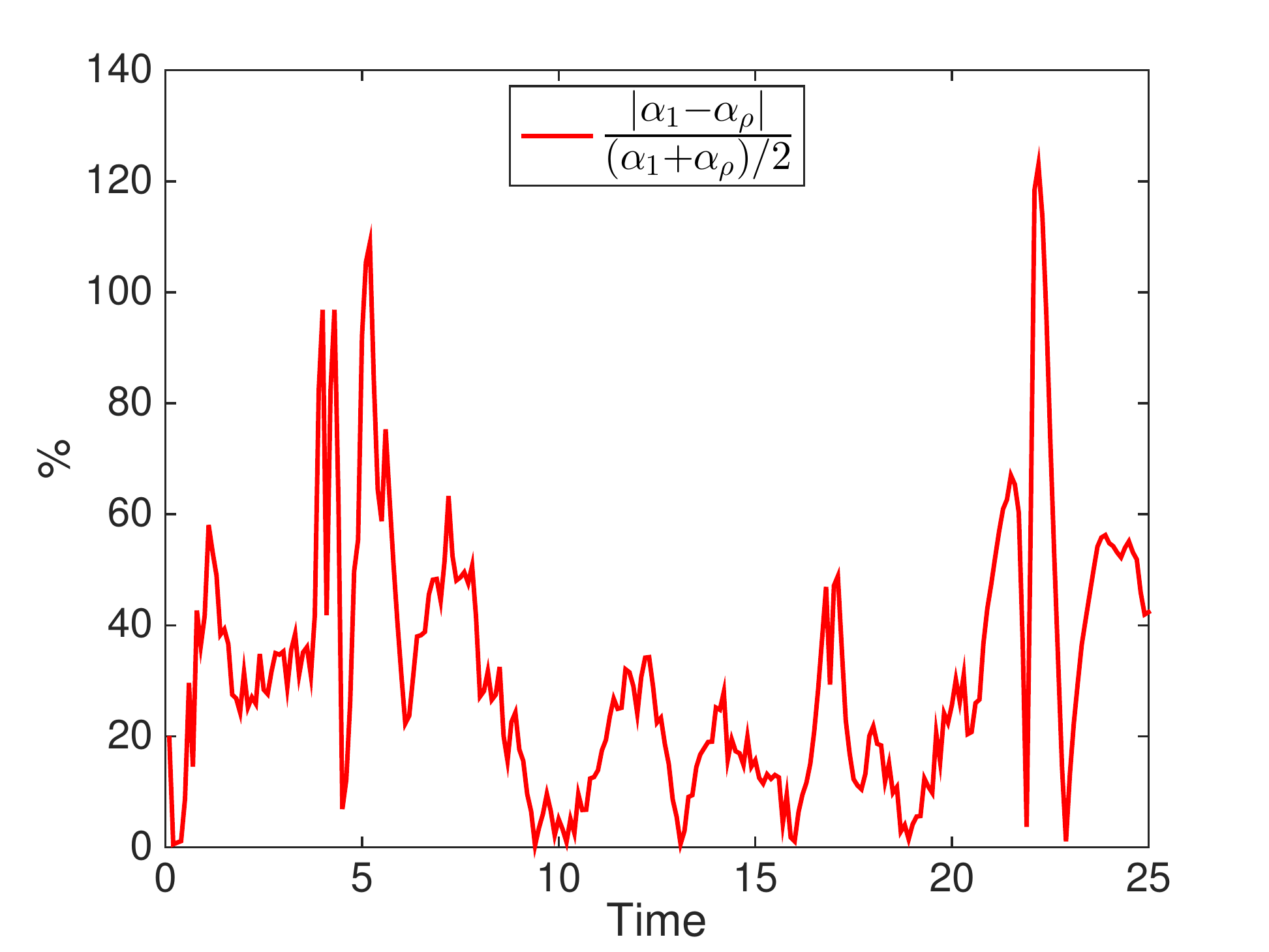}}
\end{subfigure} 
\caption{(a) Time evolution of the quantities $\alpha_{1}$ and $\alpha_{\rho}$. (b) Time evolution of the relative difference $\frac{|\alpha_{\rm 1} - \alpha_\rho|}{\left(\alpha_{\rm 1}+\alpha_\rho \right)/2}$ expressed in percentage.}
\end{figure}

In figure \ref{fig:nu1_nurho-a}, we represent the time evolution of the two quantities in equations (\ref{eq:fnorm_nu1_nurho}). We can notice that, even if the order of magnitude of the two quantities is the same, consistent differences between them are present. If we consider the time evolution of the relative difference between $  \alpha_1 $ and $\alpha_{\rho}$ (figure \ref{fig:nu1_nurho-b}), we can see that relative differences up to $100\%$ arise. As a consequence, even if the order of magnitude of $\nu_{\rm 1}$ and $\nu_\rho$ appears to be the same, it is safer not to identify the two eddy viscosities in order not to risk to neglect additional terms, with respect to the traditional formulation, which can be important also when $\nu_{\rm 1}$ and $\nu_\rho$ are slightly different between each other.  

Concluding, if, despite the low correlations values, an eddy viscosity approach is preferred, the better way to implement it could be the introduction of a dynamic procedure for the determination of $\nu_{\rm 1}$ and $\nu_{\rm \rho}$ separately.

\subsection{Similarity scale approach} 
In  the framework of a similarity scale approach, we propose instead the following models for the terms $\tau(\rho u_i, u_j)$ and $\tau(\rho,u_i)$:
\begin{subequations}
\begin{align}
& \tau(\rho u_i, u_j) = c_{\rm 1} \left(\overline{\overline{\rho u_i}\wt{u}_j} - \overline{\overline{\rho u_i}}\,\overline{\wt{u_j}}\right), \\
& \tau(\rho,u_i) = c_{\rm \rho}\left( \overline{\overline{\rho u_i}} - \overline{\frho}\,\overline{\wt{u}_i}\right),
\end{align}
\label{eq:similarity_scale_model}
\end{subequations}
where it should be noticed that the second filtering operation is realized by means of the $\overline{\cdot}$ filter, rather than the Favre filter $\wt{\cdot}$, since, in this case, the unfiltered density would be necessary, which cannot be computed in a LES (see \cite{vreman:1995}). As in the eddy viscosity approach, the two constants $c_{\rm 1}$ and $c_{\rm \rho}$ can be determined employing a dynamic procedure. 

Notice that our similarity scale approach is an extension to compressible variable density flows of the conventional similarity scale approach, first proposed in \cite{bardina:1983} and successively extended to compressible flows in \cite{vreman:1995}. The conventional similarity scale approach is given by:
\begin{equation}
\frho \theta(u_i,u_j) = c \frho \left( \overline{\wt{u}_i \wt{u}_j} - \overline{\wt{u}}_i \overline{\wt{u}}_j \right),
\label{eq:sim_scale_mod_conv}
\end{equation}
where a dynamic procedure can be employed for the determination of the constant $c$. 

\begin{figure}
\centering
\includegraphics[width=0.7\textwidth]{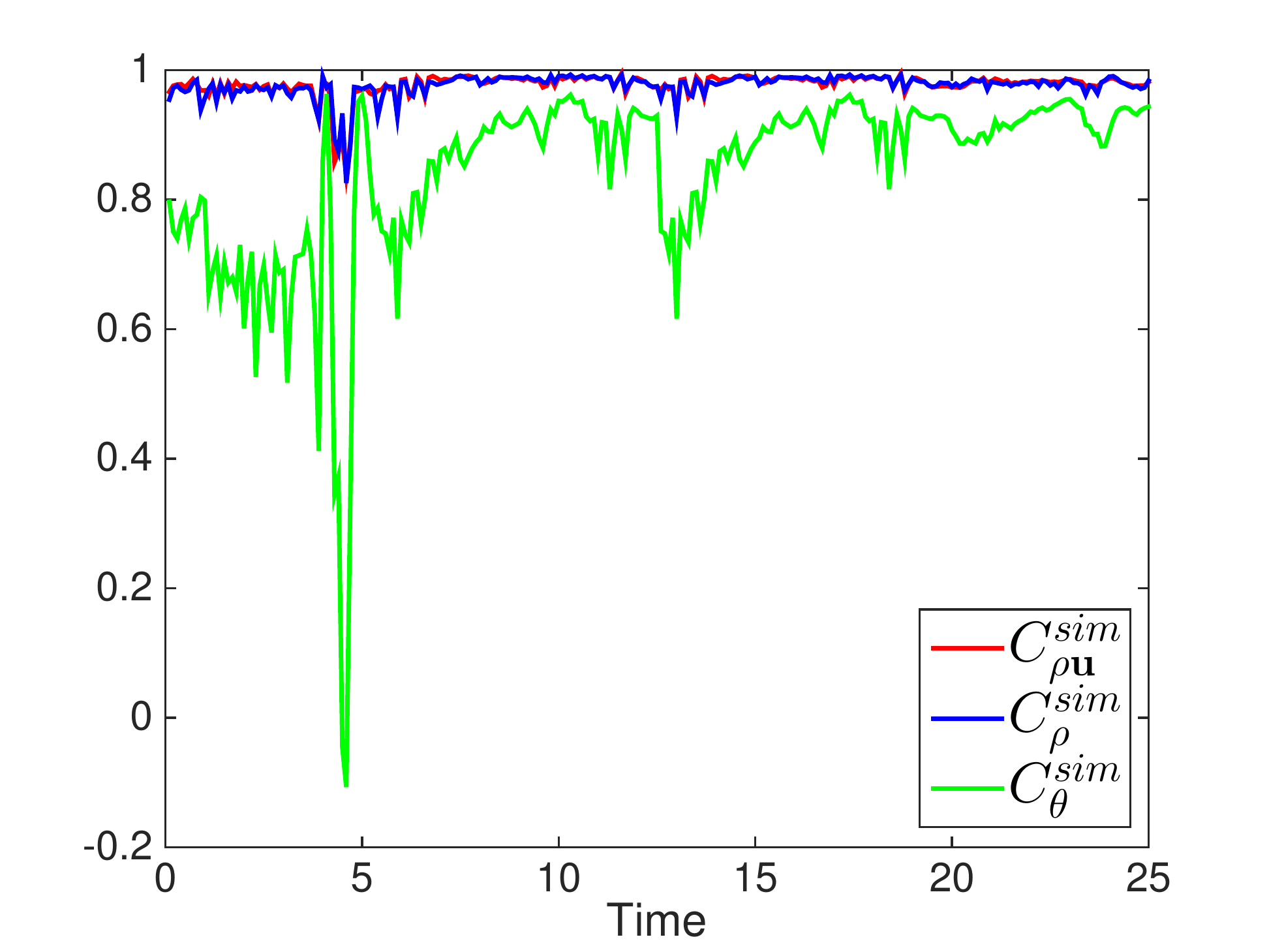}
\caption{Correlations $C_{\rho \bu}^{sim}$, $C_\rho^{sim}$ (equations (\ref{eq:corr_mom_similarity}) and  (\ref{eq:corr_similarity})) and correlation $C_\theta^{sim}$ (equation (\ref{eq:corr_theta_similarity})), as a function of time.}
\label{fig:correlations_similarity}
\end{figure}

In order to see if the introduction of similarity scale models provides better results with respect to the eddy viscosity approach, we evaluate by means of the \textit{a priori} tests the time evolution of the following correlations:
\begin{subequations}
\begin{align}
& C_{\rho \bu}^{sim} = \frac{\frac{1}{|\Omega|} \int_{\Omega} \tau(\rho u_i,u_j) \left(\overline{\overline{\rho u_i}\wt{u}_j} - \overline{\overline{\rho u_i}}\,\overline{\wt{u_j}}\right) d\bx}{\sqrt{\frac{1}{|\Omega|^2} \int_{\Omega}\left|\tau(\rho u_i,u_j)\right|^2 d\bx \int_{\Omega}\left|\left(\overline{\overline{\rho u_i}\wt{u}_j} - \overline{\overline{\rho u_i}}\,\overline{\wt{u}_j}\right)\right|^2 d\bx}}, \label{eq:corr_mom_similarity}\\
& C_{\rho}^{sim} = \frac{\frac{1}{|\Omega|} \int_{\Omega} \tau(\rho, u_i)\left( \overline{\overline{\rho u_i}} - \overline{\frho}\,\overline{\wt{u}_i}\right) d\bx}{\sqrt{\frac{1}{|\Omega|^2} \int_{\Omega}\left|\tau(\rho, u_i)\right|^2 d\bx \int_{\Omega}\left|\left( \overline{\overline{\rho u_i}} - \overline{\frho}\,\overline{\wt{u}_i}\right)\right|^2 d\bx}}. \label{eq:corr_similarity}
\end{align}
\label{eq:correlations_similarity}
\end{subequations}
These correlations are analogous to those defined in equations (\ref{eq:correlations}) for the eddy viscosity case. 
In figure \ref{fig:correlations_similarity}, the time evolution of the two correlations (\ref{eq:correlations_similarity}) is presented, together with the time evolution of the correlation $C_\theta^{sim}$ (associated to the conventional similarity scale model in equation (\ref{eq:sim_scale_mod_conv})), which is computed as follows:
\begin{equation}
C_\theta^{sim} = \frac{\frac{1}{|\Omega|} \int_{\Omega} \frho \theta(u_i,u_j)\frho\left(\overline{\wt{u}_i \wt{u}_j} - \overline{\wt{u}}_i \overline{\wt{u}}_j \right) d\bx}{\sqrt{\frac{1}{|\Omega|^2} \int_{\Omega}\left|\frho \theta(u_i,u_j) \right|^2 d\bx \int_{\Omega}\left|\frho\left(\overline{\wt{u}_i \wt{u}_j} - \overline{\wt{u}}_i \overline{\wt{u}}_j \right)\right|^2 d\bx}}.
\label{eq:corr_theta_similarity}
\end{equation} 
We can notice that the correlation values $C_{\rho \bu}^{sim}$ and $C_\rho^{sim}$ (red and blue curves) are considerably higher with respect to the values of $C_{\rho \bu}$ and $C_\rho$ obtained with the eddy viscosity approach (see figure \ref{fig:correlations}). Moreover, they are also higher with respect to $C_\theta^{sim}$ (green curve), associated to the traditional similarity scale approach. We can then conclude that a similarity scale approach as in equations (\ref{eq:similarity_scale_model}), with the dynamic computation of the two constants $c_1$ and $c_\rho$, or even a mixed model (if too little dissipation is introduced by the scale similarity model alone), could be a better choice with respect to an eddy viscosity approach and also with respect to the traditional similarity scale model for compressible flows.

\begin{figure}[]
\centering
\begin{subfigure}[\label{fig:c1_crho-a}]{
      \includegraphics[width=0.56\textwidth]{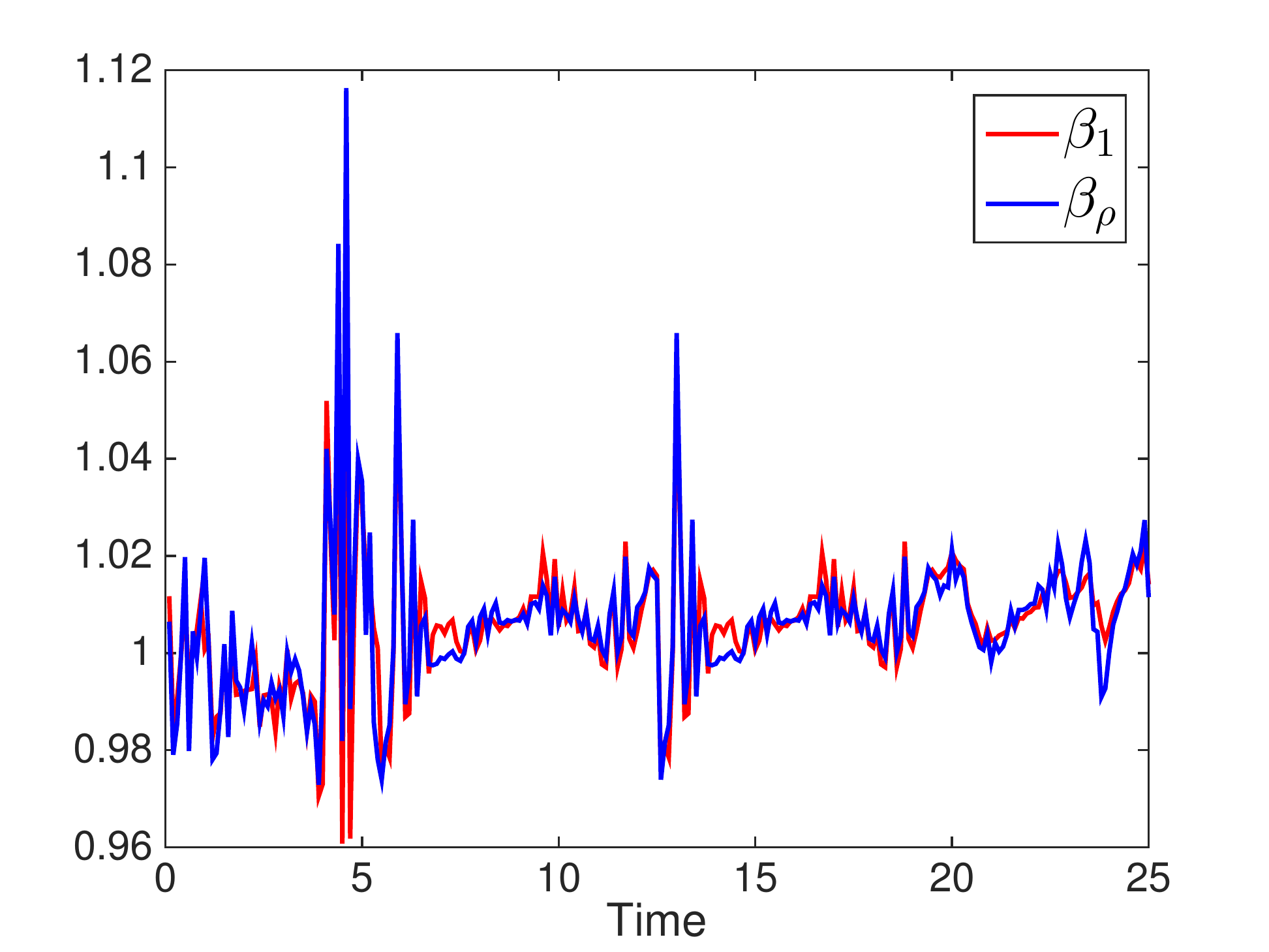}}
\end{subfigure}
\begin{subfigure}[\label{fig:c1_crho-b}]{
      \includegraphics[width=0.56\textwidth]{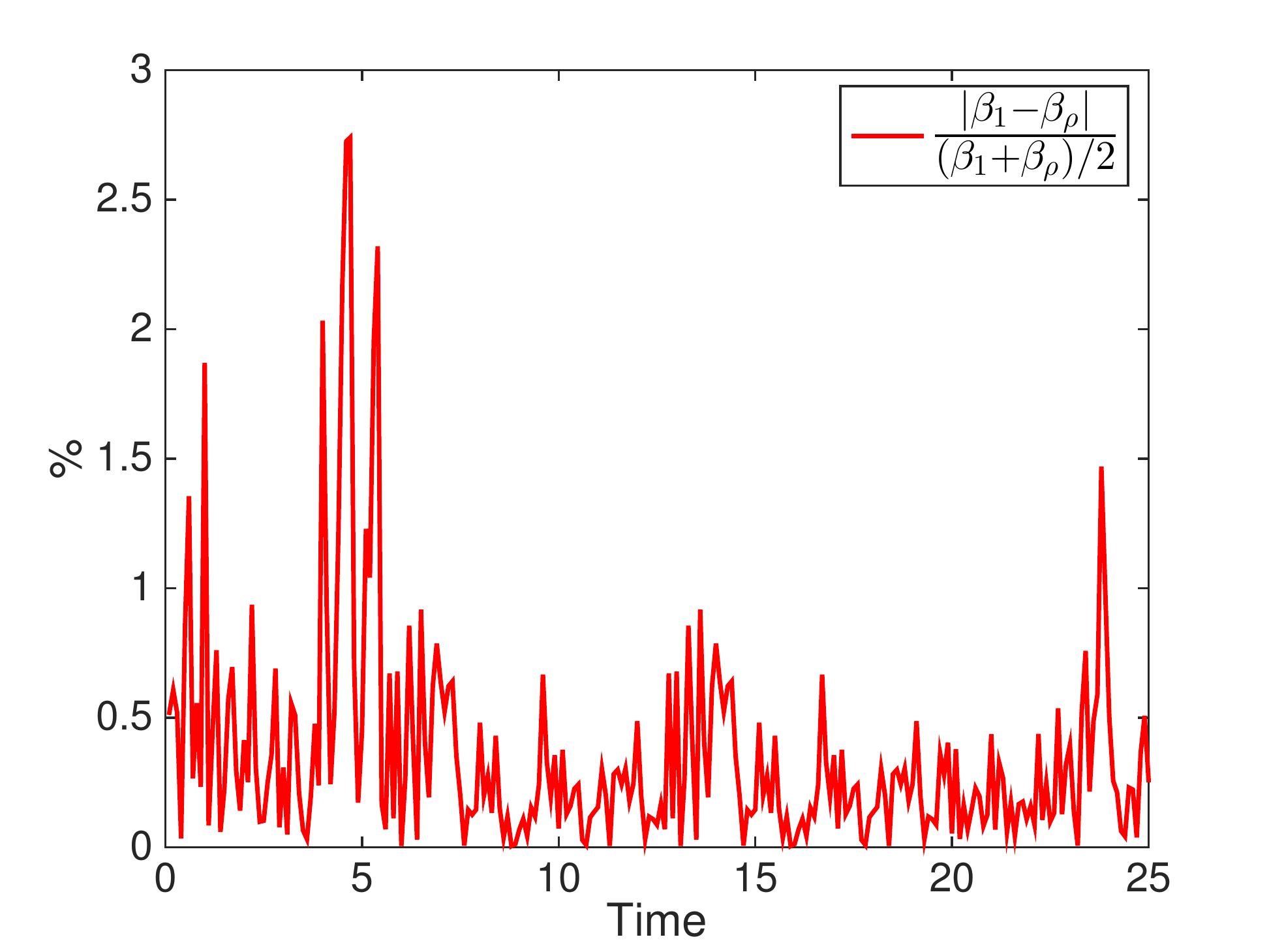}}
\end{subfigure} 
\caption{(a) Time evolution of the quantities
 $ \beta_{1} $  and  $ \beta_{\rho}. $  (b) Time evolution of the relative difference $| \beta_{\rm 1}  -\beta_{\rm \rho} | /{\left(\beta_{\rm 1} + \beta_{\rm \rho}\right)/2}$ expressed in percentage.}
\end{figure}

Analogously to what has been done for the eddy viscosity approach, we estimate the quantities $c_{\rm 1}$ and $c_{\rm \rho},$ in order to have an idea of their order of magnitude and to see if the simplification $c_{\rm 1}=c_{\rm \rho}$ can be introduced. In figure \ref{fig:c1_crho-a} we represent the time evolution of the following quantities:
\begin{equation}
 \beta_{1}   = \frac{\Vert \tau(\rho u_i,u_j) \Vert_F}{\Vert \overline{\overline{\rho u_i} \wt{u}_j} - \overline{\overline{\rho u_i}}\,\overline{\wt{u}}_j \Vert_F}, \ \ \ \ 
\beta_{\rho}  =  \frac{\Vert \tau(\rho, u_i) \Vert_F}{\Vert \overline{\overline{\rho u_i}} - \overline{\overline{\rho}}\,\overline{\wt{u}}_i \Vert_F}, 
\label{eq:fnorm_c1_crho}
\end{equation}
which are analogous to the quantities computed in equations (\ref{eq:fnorm_nu1_nurho}) for the eddy viscosity approach.
We can notice that both $\beta_{1}   $ and $\beta_{\rho} $ are similar between each other and approximately equal to $1$.
In order to better quantify the difference between $c_{\rm 1}$ and $c_{\rm \rho}$, we represent in figure \ref{fig:c1_crho-b} the relative difference between $ \beta_{1} $ and $\beta_{\rho}$: as we can see the fact that the two quantities are very similar between each other is confirmed with a relative difference which does not exceed a few percent. 
It appears, as a consequence, that the simplification $c_{\rm 1} = c_\rho$ is consistent with the findings
of the \textit{a priori} analysis.  
Notice that, in the similarity scale model case, the simplification $c_{\rm 1} = c_\rho$ does not lead to the traditional similarity scale model for compressible flows, contrarily to what happens for the eddy viscosity approach where setting $\nu_{\rm 1} = \nu_\rho$ leads to the traditional model $\frho \theta(u_i,u_j) = -\frho \nu_{\rm 1} \wt{\mathcal{S}}_{ij}$

\section{Conclusions and future developments} 
\label{sec:conclu}
In the present investigation, the theoretical work \cite{germano:2014} on LES  models for compressible variable density flows has been considered as a starting point for an improved modeling of subgrid scale
stresses in compressible flows with respect to the standard approaches.

A first numerical evaluation of the proposed ideas has been provided by means of two-dimensional \textit{a priori} tests. We have found that some terms introduced in \cite{germano:2014}, which are usually neglected in the common density weighting approach to turbulence models for compressible turbulence, are indeed not negligible. We have also found out that the modeling proposal made in \cite{germano:2014} is partially in contrast with the \textit{a priori} tests results themselves.

As a consequence, we have tried to develop alternative proposals for turbulence modelling in variable density compressible flows.
The first approach is based on the modelization of the two terms $\tau(\rho, u_i)$ and $\tau(\rho u_i, u_j)$ following the eddy viscosity hypothesis. The \textit{a priori} tests show low values for the correlations between the exact subgrid scale terms and the modeled ones, suggesting that the eddy viscosity approach may not be the better choice. However, as already noticed in \cite{abba:2001},  such low correlations are rather typical for eddy viscosity models. If, in spite the low correlations values, the eddy viscosity approach is preferred, the \textit{a priori} tests results suggest the introduction of two different,
dynamically computed eddy viscosities $\nu_{\rm 1}$ and $\nu_{\rm \rho}.$  

As expected (see \cite{vreman:1995}), the introduction of a scale similarity model for both $\tau(\rho,u_i)$ and $\tau(\rho u_i, u_j)$ considerably improves the results in terms of correlations. The correlations associated to the proposed similarity scale model are also higher than the correlations associated to the traditional similarity scale approach for compressible flows. 
Considering the \textit{a priori} tests results, the use of a scale similarity model for the terms $\tau(\rho,u_i)$ and $\tau(\rho u_i,u_j)$, possibly with the simplification $c_{\rm 1} = c_{\rho}$, appears to be the best choice. 
However, a final assessment of these proposals will require   testing both the proposed eddy viscosity model and the scale similarity model in a three-dimensional LES.

 \section*{Acknowledgements} 
 This paper is part of the first author's PhD thesis work at Politecnico di Milano.
 We are happy to acknowledge the continuous help of M. Restelli and M.Tugnoli with the application of the FEMILARO code. Several comments by F. Denaro and M.V. Salvetti have also been very useful
 to improve the presentation of some results.
  The results of this research have been achieved using the 
computational resources made available   at CINECA (Italy) by the LISA high performance computing project
 {\it DECLES: Large Eddy Simulation of Density Currents and Variable Density Flows, HPL13PJ6YS}.
 
\bibliographystyle{plain}
\bibliography{apriori_arxiv}

\begin{thebibliography}{10}

\bibitem{abba:2015}
A.~Abb\`a, L.~Bonaventura, M.~Nini, and M.~Restelli.
\newblock Dynamic models for {Large Eddy Simulation} of compressible flows with
  a high order {DG} method.
\newblock {\em Computers \& Fluids}, 122:209--222, 2015.

\bibitem{abba:2001}
A.~Abb\`a, C.~Cercignani, and L.~Valdettaro.
\newblock {Analysis of Subgrid Scale Models}.
\newblock {\em Computer and Mathematics with Applications}, 46:521--535, 2003.

\bibitem{bardina:1983}
J.~Bardina, J.H. Ferziger, and W.C. Reynolds.
\newblock Improved turbulence models based on {LES} of homogeneous
  incompressible turbulent flows.
\newblock Technical Report TF-19, Department of Mechanical Engineering,
  Stanford, 1983.

\bibitem{bassi:2017}
C.~Bassi, A.~Abb\`a, L.~Bonaventura, and L.~Valdettaro.
\newblock {Large Eddy Simulation} of gravity currents with a high-order {DG}
  method.
\newblock {\em Communications in Applied and Industrial Mathematics},
  8:128--148, 2017.

\bibitem{bassi:1997}
F.~Bassi and S.~Rebay.
\newblock {High Order Accurate Discontinuous Finite Element Method for the
  Numerical Solution of the Compressible Navier-Stokes Equations}.
\newblock {\em Journal of Computational Physics}, 131:267--279, 1997.

\bibitem{birman:2005}
V.~K. Birman, J.E. Martin, and E.~Meiburg.
\newblock The non-{Boussinesq} {Lock}-exchange problem. {Part} 2.
  {High}-resolution simulations.
\newblock {\em Journal of Fluid Mechanics}, 537:125--144, 2005.

\bibitem{bonometti:2011}
T.~Bonometti, M.~Ungarish, and S.~Balachandar.
\newblock A numerical investigation of constant volume non-{Boussinesq} gravity
  currents in deep ambient.
\newblock {\em Journal of Fluid Mechanics}, 673:574--602, 2011.

\bibitem{britter:1978}
R.E. Britter and J.E. Simpson.
\newblock Experiments on the dynamics of a gravity current head.
\newblock {\em Journal of Fluid Mechanics}, 88:223--240, 1978.

\bibitem{chavent:1989}
G.~Chavent and B.~Cockburn.
\newblock The local projection $p^0-p^1$ discontinuous {Galerkin} finite
  element method for scalar conservation laws.
\newblock {\em Mathematical Modelling and Numerical Analysis}, 23:565--592,
  1989.

\bibitem{cockburn:1989}
B.~Cockburn and C.W. Shu.
\newblock {TVB Runge-Kutta} local projection discontinuous {Galerkin} finite
  element method for conservation laws {II}: general framework.
\newblock {\em Mathematics of Computation}, 186:411--435, 1989.

\bibitem{constantinescu:2014}
G.~Constantinescu.
\newblock {LES} of lock-exchange compositional gravity currents: a brief review
  of some recent results.
\newblock {\em Environmental Fluid Mechanics}, 14:295--317, 2014.

\bibitem{germano:2014}
M.~Germano, A.~Abb\`a, R.~Arina, and L.~Bonaventura.
\newblock On the extension of the eddy viscosity model to compressible flows.
\newblock {\em Physics of Fluids}, 26, 2014.

\bibitem{germano:1991}
M.~Germano, U.~Piomelli, P.~Moin, and W.H. Cabot.
\newblock {A Dynamic Subgrid-Scale Eddy Viscosity Model}.
\newblock {\em Physics of Fluids}, 3(7):1760--1765, 1991.

\bibitem{hartel:2000}
C.~H\"artel, E.~Meiburg, and F.~Necker.
\newblock Analysis and direct numerical simulation of the flow at a
  gravity-current head. {Part} 1. {Flow} topology and front speed for slip and
  no-slip boundaries.
\newblock {\em Journal of Fluid Mechanics}, 418:189--212, 2000.

\bibitem{huppert:1980}
H.~Huppert and J.E. Simpson.
\newblock The slumping of gravity currents.
\newblock {\em Journal of Fluid Mechanics}, 99:785--799, 1980.

\bibitem{keller:1991}
J.J. Keller and Y.P. Chyou.
\newblock On the hydraulic lock-exchange problem.
\newblock {\em Journal of Applied Mathematics and Physics (ZAMP)}, 42:874--910,
  1991.

\bibitem{klemp:1997}
J.B. Klemp, R.~Rotunno, and W.C. Skamarock.
\newblock On the dynamics of a gravity current in a channel.
\newblock {\em Journal of Fluid Mechanics}, 331:169--198, 1997.

\bibitem{ooi:2007}
S.K. Ooi, G.~Constantinescu, and L.J. Weber.
\newblock {2D} {Large Eddy Simulation} of {Lock}-exchange gravity current flows
  at high {Grashof} numbers.
\newblock {\em Journal of Hydraulic Engineering}, 133:1037--1047, 2007.

\bibitem{ozgokmen:2009}
T.M. \"Ozg\"okmen, T.~Iliescu, and P.F. Fischer.
\newblock {Large Eddy Simulation} of stratified mixing in a three-dimensional
  {Lock-exchange} system.
\newblock {\em Ocean Modelling}, 26:134--155, 2009.

\bibitem{ozgokmen:2007}
T.M. \"Ozg\"okmen, T.~Iliescu, P.F. Fischer, A.~Srinivasan, and J.~Duan.
\newblock {Large Eddy Simulation} of stratified mixing in two-dimensional
  dam-break problem in a rectangular enclosed domain.
\newblock {\em Ocean Modelling}, 16:106--140, 2007.

\bibitem{sagaut:2009}
P.~Sagaut.
\newblock {\em Large Eddy Simulation for Compressible Flows.}
\newblock Springer Verlag, 2009.

\bibitem{shin:2004}
J.~Shin, S.~Dalziel, and P.F. Linden.
\newblock Gravity currents produced by lock-exchange.
\newblock {\em Journal of Fluid Mechanics}, 521:1--34, 2004.

\bibitem{simpson:1997}
J.E. Simpson.
\newblock {\em Gravity currents in the environment and in the laboratory.}
\newblock Cambridge University Press, 1997.

\bibitem{smagorinsky:1965}
J.~Smagorinsky, S.~Manabe, and J.~Leith Holloway.
\newblock Numerical results from a nine-level general circulation model of the
  athmosphere.
\newblock {\em Monthly weather Review}, 93:727--768, 1965.

\bibitem{speziale:1988}
C.G. Speziale, G.~Erlebacher, T.A. Zang, and M.Y. Hussaini.
\newblock The subgrid-scale modeling of compressible turbulence.
\newblock {\em Physics of Fluids}, 31:940--942, 1988.

\bibitem{spiteri:2002}
R.J. Spiteri and S.J. Ruuth.
\newblock A new class of optimal high-order {Strong Stability Preserving} time
  discretization methods.
\newblock {\em SIAM Journal of Numerical Analysis}, 40:469--491, 2002.

\bibitem{ungarish:2005}
M.~Ungarish.
\newblock Intrusive gravity currents in stratified ambient, shallow-water
  theory and numerical results.
\newblock {\em Journal of Fluid Mechanics}, 535:287--323, 2005.

\bibitem{vreman:1995}
B.~Vreman, B.~Geurts, and H.~Kuerten.
\newblock A-priori tests of large eddy simulation.
\newblock {\em Journal of Engineering Mathematics}, 29:299--327, 1995.

\bibitem{yoshizawa:1986}
A.~Yoshizawa.
\newblock Statistical theory for compressible turbulent shear flows, with the
  application to subgrid modeling.
\newblock {\em Physics of Fluids}, 29:2152--2164, 1986.

\end{thebibliography}

\end{document}